\documentclass[conference,compsoc]{IEEEtran}
\ifCLASSOPTIONcompsoc
  \usepackage[nocompress]{cite}
\else
  \usepackage{cite}
\fi
\ifCLASSINFOpdf
\else
\fi
\usepackage{amssymb}
\usepackage{amsmath}
\usepackage{adjustbox}
\usepackage{booktabs}
\usepackage{tikz}
\usepackage{makecell}
\usepackage{tabularx} 
\usepackage{float}
\usepackage{cite}
\usepackage{paralist}
\usepackage{multirow}
\usepackage{ragged2e}
\usepackage{tikz}
\usepackage{wasysym} 
\usepackage{algpseudocode}
\usepackage{algorithm}
\usepackage{subcaption}
\usepackage[most]{tcolorbox} % Pacchetto per tcolorbox
\usepackage{hyperref} % Per i link e l'hypertext
\usepackage{xcolor} % Per i colori
\newcommand{\takeaway}[4]{%
\tcbset{beforeafter skip balanced=0.25\baselineskip} % Spaziatura sopra e sotto il box
\begin{tcolorbox}[
    enhanced,
    breakable,
    fontupper=\linespread{.93}\selectfont,
    left=2mm, right=2mm, top=1mm, bottom=1mm
    ]
    \textbf{\hypertarget{#1}{#2.} #3} #4
\end{tcolorbox}
}
 
\usepackage{pifont}
\usepackage{xspace}
\newcommand{\cmark}{\textcolor{green}{\ding{51}}} % Green checkmark
\newcommand{\xmark}{\textcolor{red}{\ding{55}}}   % Red X

\usepackage{ntheorem}
\usepackage{framed}

\theoremstyle{break}
\theoremheaderfont{\normalfont\bfseries}
\theorembodyfont{\normalfont \itshape}
\theoremseparator{}

\definecolor{figGreen}{HTML}{2CA02C}
\definecolor{figRed}{HTML}{D62728}

\usepackage[section,numberedsection=autolabel,nogroupskip,nonumberlist]{glossaries}
\makeglossaries
\glsdisablehyper

\newacronym{fl}{FL}{Federated Learning} 
\newacronym{gia}{GIA}{Gradient Inversion Attack}
\newacronym{bn}{BN}{Batch Normalization}
\newacronym{fedsgd}{FedSGD}{Federated Stochastic Gradient Descent}
\newacronym{fedavg}{FedAVG}{Federated Averaging}
\newacronym{gan}{GAN}{Generative Adversarial Network}
\newacronym{fc}{FC}{Fully Connected}
\newacronym{cnn}{CNN}{Convolutional Neural Network}
\newacronym{dp}{DP}{Differential Privacy}
\newacronym{sa}{SA}{Secure Aggregation}
\newacronym{cdp}{CDP}{Centralized Differential Privacy}
\newacronym{ldp}{LDP}{Local Differential Privacy}
\newacronym{cml}{CML}{Centralized Machine Learning}
\newacronym{ood}{OOD}{Out-Of-Distribution}
\newacronym{ssim}{SSIM}{Structural Similarity}
\newacronym{psnr}{PSNR}{Peak Signal-to-Noise Ratio}
\newacronym{lpips}{LPIPS}{Learned Perceptual Image Patch Similarity}
\newacronym{mse}{MSE}{Mean Square Error}
\newacronym{he}{HE}{Homomorphic Encryption}
\newacronym{fhe}{FHE}{Fully Homomorphic Encryption}
\newacronym{fu}{FU}{Federated Unlearning}
\newacronym{ml}{ML}{Machine Learning}
\newacronym{niid}{non-IID}{non-Independent and Identically Distributed}
\newacronym{iid}{IID}{Independent and Identically Distributed}
\newacronym{vlm}{VLM}{Vision-Language Model}
\newacronym{mia}{MIA}{Membership Inference Attack}
\newacronym{vit}{ViT}{Vision Transformer}
\newacronym{cdf}{CDF}{Cumulative Distribution Function}
\newacronym{dsnr}{D-SNR}{Disaggregation Signal-to-Noise Ratio}
\newacronym{lda}{LDA}{Latent Dirichlet Allocation}
\newacronym{tpr}{TPR}{True Positive Rate}
\newacronym{fpr}{FPR}{False Positive Rate}

\newcommand{\giu}[1]{\textcolor{black}{#1}}

\begin{document}
\title{On the Detectability of Active Gradient Inversion Attacks in Federated Learning}



% compsoc conferences compared to traditional conferences), use this
%Telephone: (800) 555--1212, Fax: (888) 555--1212}

% \author{%
%   \IEEEauthorblockN{Anonymous Author(s)}
%   % \IEEEauthorblockA{Affiliation withheld for double-blind review}
% }

\IEEEoverridecommandlockouts 
\author{
    \IEEEauthorblockN{
        Vincenzo Carletti,
        Pasquale Foggia,
        Carlo Mazzocca,
        Giuseppe Parrella\textsuperscript{*}, 
        and Mario Vento
    }
    \IEEEauthorblockA{
        Department of Computer Information and Electrical Engineering and Applied Mathematics \\
        University of Salerno \\
        \{vcarletti, pfoggia, cmazzocca, gparrella, mvento\}@unisa.it
    }
    \thanks{\textsuperscript{*}Corresponding Author}
}


\maketitle


\begin{abstract}
One of the key advantages of Federated Learning (FL) is its ability to collaboratively train a Machine Learning (ML) model while keeping clients’ data on-site. However, this can create a false sense of security. Despite not sharing private data increases the overall privacy, prior studies have shown that gradients exchanged during the FL training remain vulnerable to Gradient Inversion Attacks (GIAs). These attacks allow reconstructing the clients' local data, breaking the privacy promise of FL. 
GIAs can be launched by either a passive or an \emph{active} server. In the latter case, a malicious server manipulates the global model to facilitate data reconstruction. While effective, earlier attacks falling under this category have been demonstrated to be detectable by clients, limiting their real-world applicability. 
Recently, novel active GIAs have emerged, claiming to be far stealthier than previous approaches. 
This work provides the first comprehensive analysis of these claims, investigating four state-of-the-art GIAs.
We propose novel lightweight client-side detection techniques, based on statistically improbable weight structures and anomalous loss and gradient dynamics. Extensive evaluation across several configurations demonstrates that our methods enable clients to effectively detect active GIAs without any modifications to the FL training protocol.
\end{abstract}

% \fi
\IEEEpeerreviewmaketitle

\section{Introduction}\label{sec:intro}

\gls{ml} models rely heavily on user data to learn meaningful patterns and deliver accurate predictions. However, growing awareness of data privacy risks has made individuals increasingly hesitant to share personal information, highlighting the need for privacy-preserving approaches to train \gls{ml} models~\cite{privacy_risk}. \gls{fl}~\cite{mcmahan2017communication} addresses these concerns by enabling multiple clients to collaboratively train a global model on a shared task without directly exposing their local data \cite{bellavista2021decentralised}. In each training round, clients compute model updates (i.e., gradients or weight differences) based on their private data. These updates are then transmitted to a central server, which aggregates them to update the global model. Given these properties, \gls{fl} achieves a higher privacy level than traditional centralized approaches. Unfortunately, model updates can still be exploited to gain information on the training data, breaking the users' privacy. 

\begin{figure}[ht]
    \centering
    \includegraphics[width=1.0\columnwidth]{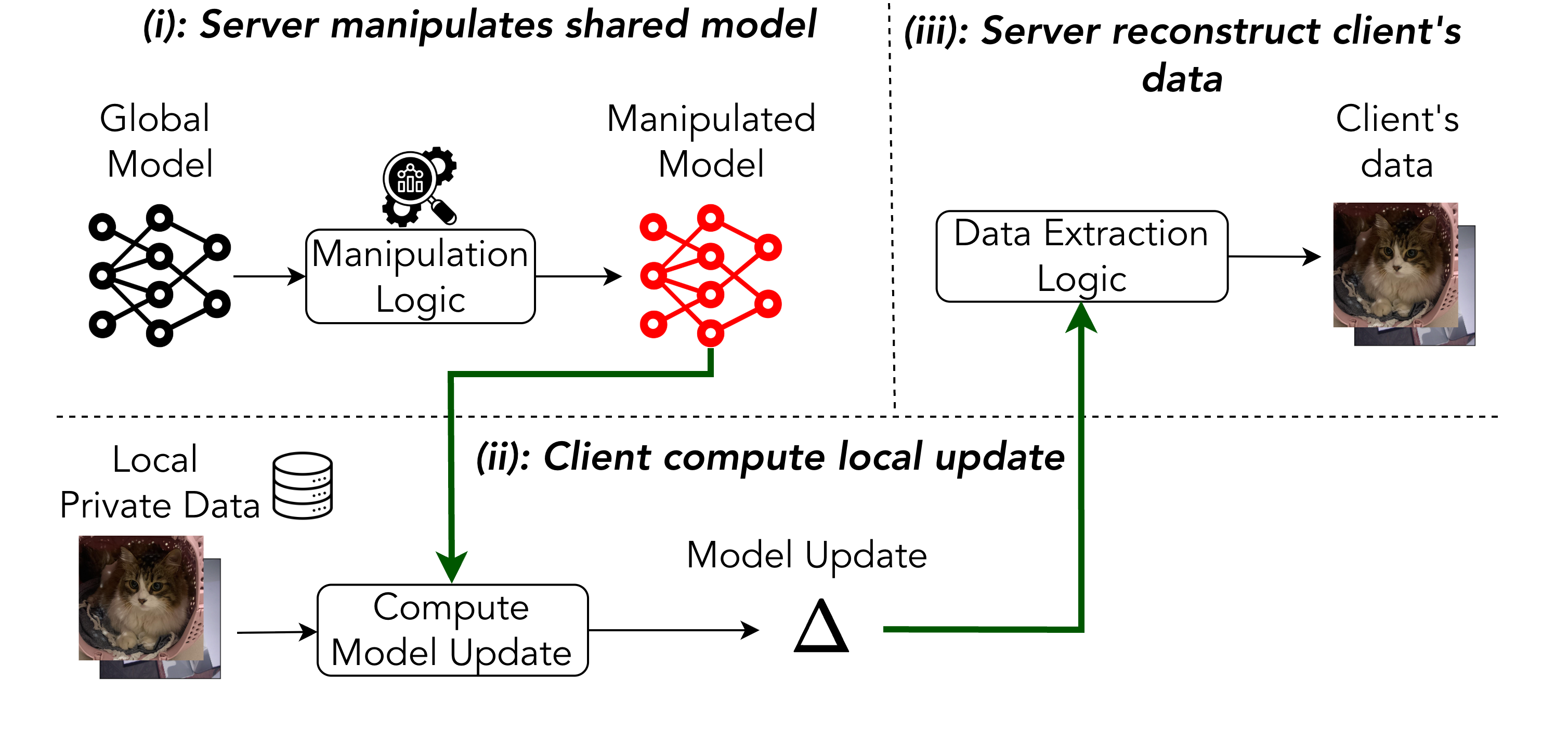}
    \caption{Graphical illustration of an active GIA workflow. \textit{(i)} The server manipulates the global model and distributes it to clients. \textit{(ii)} Each client performs local training using the manipulated global model to compute model updates. \textit{(iii)} The server receives these updates and exploits them to reconstruct the client's training data.}    
    \label{fig:active_gia_example}
\end{figure} 

A growing body of research has demonstrated that \glspl{gia} can effectively recover private data in \gls{fl} by exploiting the shared gradients~\cite{our_sok,du2024sok}. Such attacks are typically launched by a malicious server, which can operate in either a passive or an active manner. In the passive case, the server adopts an \emph{honest-but-curious} behavior: it faithfully follows the \gls{fl} protocol while covertly observing the gradients exchanged during training to infer private information about clients’ data. In contrast, in the \emph{active} case, the server deliberately manipulates the process (e.g., altering the global model) to facilitate data reconstruction. Passive attacks are inherently harder to detect since the server does not deviate from the standard \gls{fl} protocol~\cite{du2024sok,our_sok}. In contrast, active attacks can achieve greater reconstruction accuracy but are typically easier for clients to detect, which may limit their practicality in real-world deployments~\cite{du2024sok,our_sok,garov2024hiding,wang2025hear}. Figure \ref{fig:active_gia_example} illustrates the high-level workflow of an active \gls{gia}. 

\smallskip
\noindent \textbf{\emph{Active} GIAs.} 
Active \glspl{gia} involve a malicious server that employs model manipulation strategies to achieve even perfect reconstruction of client data from their model updates. Early works in this domain add carefully-crafted layers into the global model to facilitate data reconstruction~\cite{rtf,loki}. However, these architectural manipulations are easily detected by clients, as participants in an \gls{fl} system pre-agree on the model's architecture. Consequently, the focus shifted to attacks that manipulate only the model weights without altering the architecture~\cite{magnification,trap_w,trap_w_sa_ddp}. Despite this evolution, recent studies have shown that these manipulations remain detectable, often due to the distinctive patterns or simple logic they embed~\cite{wang2025hear, garov2024hiding}. 
Advanced approaches claim to offer stealthier weight manipulation techniques~\cite{mkor,Shi2023ScaleMIAAS}, yet they have not been rigorously evaluated from the standpoint of client-side detectability~\cite{our_sok}. 

More recently, a new class of data-driven model manipulation attacks has emerged, aiming to perform active \glspl{gia} without architectural manipulation and, crucially, to evade client-side detection~\cite{garov2024hiding,shan2025geminio}. Some of these methods even satisfy formal guarantees of stealthiness~\cite{garov2024hiding}, measured by the \gls{dsnr} metric introduced in the same work. However, subsequent research has demonstrated that this metric is ineffective even against earlier active \glspl{gia}~\cite{rtf} and susceptible to false positives~\cite{wang2025hear}. As a result, this \gls{dsnr} proves inadequate for identifying these newer attacks~\cite{mkor,Shi2023ScaleMIAAS,garov2024hiding,shan2025geminio}, leaving the true threat they pose to realistic \gls{fl} systems largely unexamined~\cite{our_sok}.

\smallskip
\noindent \textbf{Contributions.} The main contributions of this work can be summarized as follows. 
\begin{itemize}
    \item We provide the first comprehensive analysis of four state-of-the-art active \glspl{gia}, systematically evaluating their attack mechanisms and their claims of stealthiness.
    \item We propose a suite of lightweight, client-side detection techniques that require no modifications to the \gls{fl} training protocol. Our methods identify key attack artifacts by focusing on statistically improbable weight structures and anomalous loss and gradient dynamics.
    \item We conduct an extensive experimental evaluation across several \gls{fl} configurations. Experimental findings demonstrate that our methods enable clients to effectively detect these advanced active \glspl{gia}. 
\end{itemize}

\section{Background \& Related Work}\label{sec:background}

\subsection{Background}\label{sec:background_sub}

\smallskip \noindent \textbf{Federated Learning.}
We consider a standard \gls{fl} deployment where a central server coordinates a group of clients to collaboratively train a global model while keeping their local data private. Our work focuses on \glspl{gia} within the horizontal \gls{fl} setting, where clients operate over a common feature space but possess disjoint data samples. This configuration is widely recognized as the reference scenario in the \gls{gia} literature~\cite{our_sok,du2024sok}. 

Let \(\mathcal{C}\) denote the set of clients in the federation. Each client \(c\in\mathcal{C}\) holds a local private dataset \(D_c=\{(x_{c,i},y_{c,i})\}_{i=1}^{N_c}\), with \(x_{c,i}\in\mathcal{X}\) and \(y_{c,i}\in\mathcal{Y}\). Given the model \(f_{\theta}\), the local empirical risk at client \(c\) is:
\begin{equation}
    f_c(\theta) = \frac{1}{N_c}\sum_{i=1}^{N_c}\mathcal{L}\big(f_{\theta}(x_{c,i}),\,y_{c,i}\big),
\end{equation}
while the global \gls{fl} objective is:
\begin{equation}
    \min_{\theta} f(\theta) \;=\; \frac{1}{|\mathcal{C}|}\sum_{c\in\mathcal{C}} f_c(\theta).
\end{equation}

Training proceeds over multiple rounds until convergence. In each round \(t\), the server selects a subset of clients \(\mathcal{K}\subset\mathcal{C}\). Each client receives the current global model parameters \(\theta_t\) and performs local optimization for \(E\) epochs using mini‑batches of size \(B\). The specific client update mechanism and the server-side aggregation rule are determined by the chosen \gls{fl} algorithm. In \gls{fedsgd}~\cite{mcmahan2017communication}, each client computes an aggregate gradient over a sampled mini‑batch and transmits it to the server. Then, the server updates the global model by averaging the received gradients:
\begin{equation}
    \theta_{t+1} \gets \theta_t - \eta\frac{1}{|\mathcal{K}|}\sum_{c\in\mathcal{K}} g_c,
\end{equation}
where \(g_c=\sum_{i=1}^{B}\nabla_{\theta}\mathcal{L}(f_{\theta_t}(x_{c,i}),y_{c,i})\) and \(\eta\) denotes the global learning rate.

\gls{fedavg}~\cite{mcmahan2017communication} is widely recognized as the reference baseline in \gls{fl}. It mitigates communication overhead by allowing clients to perform multiple local updates before transmitting their models to the server. Specifically, each client \(c\in\mathcal{K}\) initializes its local model as \(\theta_c^{(t,0)}=\theta_t\), performs \(\tau\) steps of local gradient descent, and uploads the final model \(\theta_c^{(t,\tau)}\). The server then aggregates the received local models, optionally weighted by \(N_c\), to obtain the next global model:
\begin{equation}
    \theta_{t+1} \gets \frac{1}{|\mathcal{K}|}\sum_{c\in\mathcal{K}} \theta_c^{(t,\tau)}.
\end{equation}

\smallskip \noindent \textbf{Gradient Inversion Attacks.}
\glspl{gia} represent a significant privacy threat in \gls{fl}, where adversaries attempt to reconstruct private 
training data from the gradients or model updates shared during the training process~\cite{dlg,invgrad,our_sok,du2024sok}. 
In the \gls{gia} literature, the primary adversary is typically assumed to be the central server, which leverages its access to both the shared model and the model updates exchanged by clients (as well as available auxiliary information) in an attempt to reconstruct the clients' private data~\cite{our_sok,du2024sok}. 
A common approach used to categorize \glspl{gia} is based on the underlying threat model, particularly with respect to the capabilities granted to the central server~\cite{our_sok,du2024sok}. Under this perspective, \glspl{gia} are broadly classified as \emph{passive} or \emph{active} \glspl{gia}.
In passive \glspl{gia}, the server is assumed to be honest-but-curious; it may utilize auxiliary information to improve input reconstruction, but it does not interfere with the standard \gls{fl} protocol. 
In contrast, active \glspl{gia} assume a malicious server that is permitted to manipulate or deviate from the standard \gls{fl} training procedure to facilitate data reconstruction. 
\subsection{Related Work}\label{sec:related_works}

\begin{table*}[h!]
\centering
\small
\renewcommand{\arraystretch}{1.8} % Increase row spacing
\small
\begin{adjustbox}{max width=\textwidth}
\begin{tabular}{c >{\raggedright\arraybackslash}p{2.8cm} c c c >{\arraybackslash}p{8cm}}
\toprule
\shortstack{\textbf{Category}\\\textbf{ }} & \shortstack{\textbf{Attack Method}\\\textbf{ }} & \shortstack{\textbf{Architectural}\\\textbf{Manipulations}} & \shortstack{\textbf{Previous}\\\textbf{Works}} & \shortstack{\textbf{Our Work}\\\textbf{ }} & \shortstack{\textbf{Attack Logic}\\\textbf{ }} \\
\midrule

\multirow{10}{*}{\shortstack{Handcrafted\\GIAs}} & Fowl et al.~\cite{rtf} (ICLR'22) & \cmark & \CIRCLE & \Circle & Inserts imprinting modules containing linear layers near input to capture raw data features through gradient manipulation. \\

& Wen et al.~\cite{magnification} (ICML'22) & \xmark & \CIRCLE & \Circle & Zeros out most weights in final classification layer to amplify gradient contributions from individual training examples. \\

& Boenisch et al.~\cite{trap_w} (EuroS\&P'23) & \xmark & \CIRCLE & \Circle & Designs \textit{trap weights} in linear layers to precisely control ReLU activations for single data point gradient isolation. \\

& Zhao et al.~\cite{loki} (S\&P'24) & \cmark & \CIRCLE & \Circle & Inserts client-specific convolutional kernels and FC layers to maintain distinguishable weight gradients under secure aggregation. \\

& Wang et al.~\cite{mkor} (WACV'24) & \xmark & \Circle & \CIRCLE & Introduces subtle parameter modifications to increase orthogonality between prior knowledge and gradients while preserving architecture. \\

& Shi et al.~\cite{Shi2023ScaleMIAAS} (NDSS'25) & \xmark & \Circle & \CIRCLE & Employs two-phase attack combining efficient latent representation reconstruction with generative decoding for large-scale recovery. \\

\hline

\multirow{3}{*}{\shortstack{Learned\\GIAs}} & Garov et al.~\cite{garov2024hiding} (ICLR'24) & \xmark & \LEFTcircle & \CIRCLE & Jointly trains secret decoder with shared model using data-driven optimization to satisfy formal stealthiness requirements. \\

& Shan et al.~\cite{shan2025geminio} (ICCV'25) & \xmark & \Circle & \CIRCLE & Leverages Vision-Language Models to enable semantically targeted attacks with natural language specification of valuable data types. \\

\bottomrule
\end{tabular}
\end{adjustbox}
\renewcommand{\arraystretch}{1}
\caption{Overview of Active GIAs detectability in existing literature. Symbols: \CIRCLE = included/detected; \LEFTcircle = partially included/detected; \Circle = not included/not detected.}
\label{tab:active_gia_detectability}
\end{table*}

\smallskip \noindent \textbf{Active Gradient Inversion Attacks.}
Active \glspl{gia} can be further classified according to the server-side attack logic into two categories: \textit{handcrafted \glspl{gia}} and \textit{learned \glspl{gia}}. In handcrafted \glspl{gia}, adversaries explicitly modify specific global model's layers or parameters following predefined strategies based on analytical insights into the gradient computation process. In contrast, learned \glspl{gia} employ data-driven optimization strategies to automatically discover the optimal global model's manipulations, learning to maximize data extraction effectiveness without requiring manual design of model layers or parameters. Table~\ref{tab:active_gia_detectability} presents an overview of active \glspl{gia} and their detectability analysis in existing literature. 

Early work on active \glspl{gia} typically introduced handcrafted manipulations to the global model. Fowl et al.~\cite{rtf} proposed an \textit{imprinting module}, inserting two linear \gls{fc} layers near the input to capture raw data using manipulated weights and biases. However, this insertion creates easily detectable architectural anomalies, such as anomalous parameter counts or layer patterns~\cite{garov2024hiding,wang2025hear}. Wen et al.~\cite{magnification} introduced gradient magnification, which iteratively zeroes out most weights in the final classification layer. This amplifies a single sample's gradient contribution, isolating it from the batch average for reconstruction. Yet, this attack also introduces detectable irregularities in the parameter distribution~\cite{garov2024hiding,wang2025hear}. Boenisch et al.~\cite{trap_w,trap_w_sa_ddp} used \textit{trap weights} to control neuron activations in the first linear layer, ensuring only one data point contributes to specific gradients for analytical recovery. For common \glspl{cnn}, this method necessitates modifying preceding layers to propagate the input, creating atypical and easily detected weight patterns~\cite{garov2024hiding,wang2025hear}.
Zhao et al.~\cite{loki} insert a malicious module, comprising customized convolutional kernels and linear \gls{fc} layers, into the global model~\cite{loki}. This design makes gradients client-identifiable, bypassing \gls{sa} protocols and enabling the server to analytically reconstruct private data samples~\cite{loki}. However, these overt architectural modifications are susceptible to client-side detection~\cite{garov2024hiding,wang2025hear}.

Within the category of handcrafted \glspl{gia}, two works have proposed more sophisticated attack strategies whose detectability remains overlooked~\cite{our_sok}. Wang et al.~\cite{mkor} introduced subtle parameter modifications aimed at increasing the orthogonality between prior knowledge and gradients, while maintaining the original model architecture. Shi et al.~\cite{Shi2023ScaleMIAAS} presented a two-phase attack that combines efficient latent representation reconstruction with generative decoding, enabling data recovery even under \gls{sa}.

Recently, learned \glspl{gia} have employed sophisticated strategies to craft the global model for data extraction. These include the Secret Embedding and REconstruction (SEER) framework, which jointly trains a secret decoder with the global model to enable stealthy data reconstruction from large batches~\cite{garov2024hiding}. Other recent approaches leverage \gls{vlm} to facilitate semantically targeted attacks, allowing the adversary to optimize the malicious model to reconstruct high-value samples specified using natural language queries~\cite{shan2025geminio}.

\smallskip \noindent \textbf{Detectability of Gradient Inversion Attacks.}
The detectability of active \glspl{gia} has received limited attention in the literature. This is largely because early works on active \glspl{gia} introduced model manipulations that are easily detectable, as acknowledged in prior studies~\cite{rtf,trap_w,our_sok,du2024sok,wang2025hear,garov2024hiding,Shi2023ScaleMIAAS}. 
\giu{
Garov et al.~\cite{garov2024hiding} proposed the \gls{dsnr} metric, which identifies some \glspl{gia}~\cite{magnification} by detecting dominant gradient components. However, this metric is ineffective against attacks that do not produce such signals, and is unable to detect even earlier \glspl{gia} that introduce easily detected model manipulation~\cite{wang2025hear}. To date, the most comprehensive analysis of detectability is provided by Wang et al.~\cite{wang2025hear}. However, their work consists primarily of a theoretical analysis of the SEER attack. They posit that SEER's core mechanism is critically dependent on \gls{bn} layers, which are frequently deprecated in modern architectures and known to destabilize \gls{fl} training~\cite{wang2025hear}. While Wang et al.~\cite{wang2025hear} also propose a complementary Z-score-based detector, its practical applicability is limited, as it relies on monitoring a predefined, static set of "vulnerable" layers for statistical anomalies~\cite{wang2025hear}. To address these limitations, we propose a novel detection logic that is model- and dataset-agnostic. This generalizable design, extendable to future threats, enables the first client-side detectability analysis of state-of-the-art active \glspl{gia}~\cite{mkor,Shi2023ScaleMIAAS,shan2025geminio}.
}

\section{Threat Model}\label{sec:threat_model}

This work adopts a threat model in which the central server is assumed to be both \emph{active} and \emph{malicious}. This threat model aligns with those employed in \glspl{gia} literature~\cite{trap_w,trap_w_sa_ddp,magnification,rtf,loki,shan2025geminio,garov2024hiding,mkor,our_sok,du2024sok}.

\smallskip
\noindent \textbf{Attacker.}
The server can actively interfere with the \gls{fl} training process by manipulating the global model parameters. To strengthen the effectiveness of its attacks, the server may also utilize an auxiliary dataset $D_{aux}$, which resembles but does neither partially overlap with the clients' training data, a widely accepted assumption in \gls{gia} literature~\cite{our_sok,du2024sok}. 

\smallskip
\noindent \textbf{Defender.}
In our analysis, we adopt the perspective of a client who receives a potentially compromised global model from the server and seeks to determine whether the model is legitimate or has been maliciously tampered with for a \glspl{gia}. We assume that the client is aware of the model architecture, possesses private local data, and is capable of analyzing both the received model and the computed model updates for evidence of manipulation. Additionally, we assume that each client retains its own locally trained model from the most recent \gls{fl} round in which it participated. This requirement imposes a minimal memory overhead, as it necessitates storing a single additional copy of the model parameters.
This client-centric viewpoint is crucial for understanding the practical security and privacy implications of these attacks in real-world \gls{fl} deployments, where participants must be able to verify the integrity of the global models to preserve trust in the \gls{fl} process.
\begin{figure}[t]
    \centering
    \begin{adjustbox}{max width=\columnwidth}
        \includegraphics{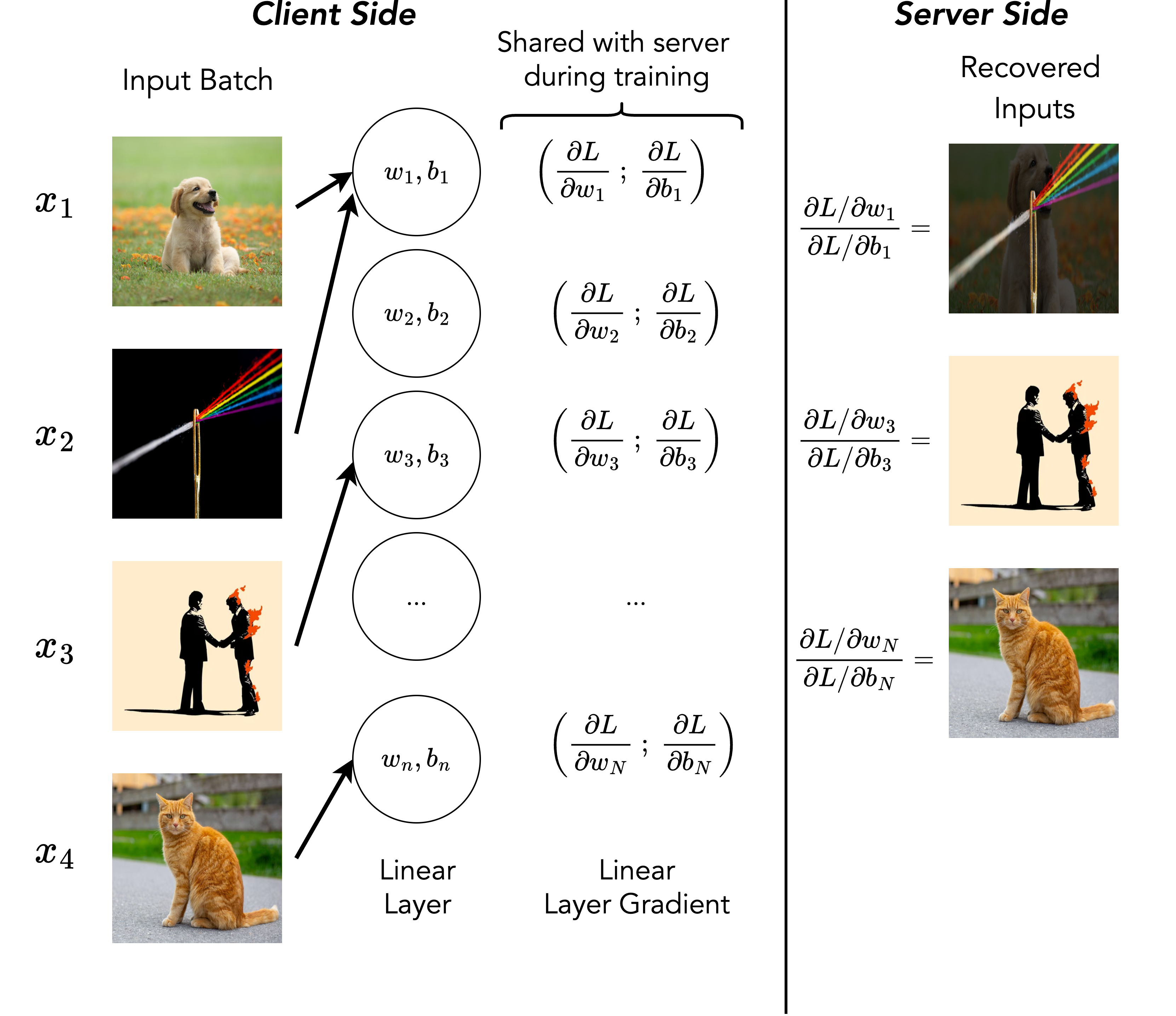}
    \end{adjustbox}
    \caption{Graphical illustration of linear layer leakage. Arrows indicate that each input exclusively activates a specific neuron in the linear layer. When a single input is present in the batch, its value can be immediately reconstructed from the corresponding gradients. However, when multiple inputs activate different neurons simultaneously, the resulting gradients represent a weighted sum of those inputs, making recovery more complex~\cite{trap_w}.}    \label{fig:linear_leakage_example}
\end{figure} 

\section{Analysis of Handcrafted GIAs} \label{sec:analysis_handcrafted_gias}
Handcrafted \glspl{gia} are analytical attack strategies where an adversary explicitly designs a parameter transformation function $\mathcal{T}: \Theta \to \Theta$, mapping the original parameter space $\Theta$ of a neural network $f_{\theta}$ to a malicious one. Given benign model parameters $\theta$, the adversary $\mathcal{A}$ (i.e., the server) crafts the malicious parameters $\theta^{\mathcal{A}}$ as follows:
\begin{equation}\label{eq:handcrafted_gias}
\theta^{\mathcal{A}} = \mathcal{T}(\theta; \mathcal{R}, D_{\text{aux}})
\end{equation}
where $\mathcal{R}$ denotes a predefined set of transformation rules and $D_{aux}$ is an optional auxiliary dataset used to derive statistics for the transformation.

The malicious server distributes $\theta^{\mathcal{A}}$ to the clients, who compute gradient updates $\nabla_{\theta^{\mathcal{A}}} \mathcal{L}$ on a private batch of data $\mathcal{X} = \{x_1, \ldots, x_B\}$ using the manipulated model. The transformation $\mathcal{T}$ is designed to create a deterministic, invertible mapping between the client's inputs $\mathcal{X}$ and the aggregated gradient $\nabla_{\theta^{\mathcal{A}}} \mathcal{L}$, enabling $\mathcal{A}$ to analytically recover the private data using a reconstruction function $\mathcal{F}$, without iterative optimization:
\begin{equation}\label{eq:handcrafted_recon}
\hat{\mathcal{X}} = \mathcal{F}(\nabla_{\theta^{\mathcal{A}}} \mathcal{L})
\end{equation}
where $\hat{\mathcal{X}}$ represents the reconstructed data.
A key principle frequently leveraged to craft the transformation rules $\mathcal{R}$ is that gradients in linear layers inherently encode a weighted average of the input data~\cite{rtf,trap_w,loki}, as illustrated in Figure~\ref{fig:linear_leakage_example}. Even passive adversaries can exploit this property for partial data recovery~\cite{trap_w}. Handcrafted \glspl{gia} amplify this leakage by manipulating the model to establish a one-to-one correspondence between inputs and gradient components.

In the following, we examine two state-of-the-art attacks proposed by Wang et al.~\cite{mkor} and Shi et al.~\cite{Shi2023ScaleMIAAS}, both of which fall under this category and have not yet been assessed from a detectability perspective.

\subsection{Binning Property-based GIAs}\label{sec:binning}
The binning property attack is a handcrafted \glspl{gia} that exploits the analytical link between gradients and inputs in linear layers. By manipulating two consecutive linear layers, termed \emph{imprint modules}, the adversary creates a deterministic mapping from input samples to neuron activations, enabling perfect data reconstruction from aggregated gradients.

\smallskip
\noindent\textbf{Intuition and Attack Steps.}
The attack operates by partitioning the latent feature space into a set of discrete bins. This is accomplished by manipulating the first linear layer, defined by its weight matrix $W_1 \in \mathbb{R}^{k \times d}$ and bias vector $b_1 \in \mathbb{R}^k$. In this formulation, $d$ represents the dimensionality of the input $x \in \mathbb{R}^d$, while $k$ denotes the number of neurons in this layer. The core mechanism is designed to make each input sample $x$ uniquely identifiable by the highest-indexed neuron it activates. This process is achieved as follows:
\begin{enumerate}
    \item \textbf{Feature Selection:} Using an auxiliary dataset $D_{\text{aux}}$, the adversary selects a feature direction vector $v_{feature} \in \mathbb{R}^d$. This vector defines a scalar projection $p(x) = <v_{feature}, x>$.
    \item \textbf{Quantile Binning:} The empirical Cumulative Distribution Function (CDF) of $p(x)$ is computed over $D_{\text{aux}}$. The range of $p(x)$ is then divided into $k$ intervals using quantile boundaries $q_1 < q_2 < \ldots < q_k$.
    \item \textbf{Weight Manipulation:} All row vectors of the first layer's weight matrix, denoted $W_{1,r,:}$ for $r \in \{1, \ldots, k\}$, are set to be identical to the feature vector's transpose: $W_{1,r,:} = v_{feature}^T$.
    \item \textbf{Bias Manipulation:} The elements of the bias vector $b_1$, denoted $b_{1,r}$, are set to the negative quantile boundaries: $b_{1,r} = -q_r$. The pre-activation of the $r$-th neuron is thus $a_r = W_{1,r,:}x + b_{1,r} = p(x) - q_r$. With a ReLU activation, neuron $r$ activates only if $p(x) > q_r$. An input $x$ is uniquely identified by the neuron with the highest index $r$ that it activates.
    \item \textbf{Second Layer Uniformity:} The parameters of the second linear layer ($W_2 \in \mathbb{R}^{o \times k}$, $b_2 \in \mathbb{R}^o$) are set to uniform values to ensure a predictable gradient flow, facilitating analytical reconstruction.
\end{enumerate}
We refer to the original papers~\cite{rtf,Shi2023ScaleMIAAS} and Appendix~\ref{sec:appendix} for a full description of binning property-based attacks.

\smallskip
\noindent\textbf{Transformation Function.}
The transformation rules $\mathcal{R}_{\text{binning}}$ are formally defined as:
\begin{equation}
\mathcal{R}_{\text{binning}} = \left\{
    \begin{alignedat}{2}
        & W_{1, r, :} = v_{feature}^T  \quad &&\forall r \in \{1, \ldots, k\} \\
        & b_{1, r}    = -q_r             \quad &&\forall r \in \{1, \ldots, k\} \\
        & W_{2, i, :} = v_{uniform}^T    \quad &&\forall i \in \{1, \ldots, o\} \\
        & b_{2, i}    = b_{uniform}      \quad &&\forall i \in \{1, \ldots, o\}
    \end{alignedat}
\right.
\end{equation}
where $v_{uniform} \in \mathbb{R}^k$ is a constant vector and $b_{uniform} \in \mathbb{R}$ is a constant scalar. The complete transformation function is $\theta^\mathcal{A} = \mathcal{T}(\theta; \mathcal{R}_{\text{binning}}, D_{\text{aux}})$.

\smallskip \noindent\textbf{Reconstruction Function.}
The manipulated layer structure ensures that the aggregated gradient of the loss $\mathcal{L}$ with respect to the layer's parameters, $W_1$ and $b_1$, analytically encodes the input samples. For an input $x_p$ that uniquely activates neuron $r$ as its highest-indexed bin, its contribution can be isolated from the aggregated batch gradient. The adversary reconstructs the input $\hat{x}_p$ associated with the bin between quantiles $q_r$ and $q_{r+1}$ by applying the following closed-form function $\mathcal{F}$ to the gradients:
\begin{equation}
    \hat{x}_p^T = \mathcal{F}(\nabla \mathcal{L}) = \frac{\nabla_{W_{1,r+1,:}} \mathcal{L} - \nabla_{W_{1,r,:}} \mathcal{L}}{\nabla_{b_{1,r+1}} \mathcal{L} - \nabla_{b_{1,r}} \mathcal{L}}
\end{equation}
The subtraction operation in both the numerator and denominator effectively isolates the gradient contribution from the single input $x_p$, separating it from the contributions of all other samples within the aggregated batch.

\smallskip
\noindent\textbf{From Feature Reconstruction to Input Recovery.}
A critical limitation of this attack strategy is that it reconstructs the immediate input to the manipulated linear layers. In standard \gls{cnn} architectures, these layers are not located near the network's input; rather, they are positioned deep within the model, typically comprising the final classification head. A graphical high-level overview of a \gls{cnn} architecture can be found in Figure~\ref{fig:model_architecture_example} in Appendix~\ref{sec:appendix_background}. Consequently, these layers operate on high-level Latent Space Representations (LSRs) that are processed by the preceding convolutional backbone, rather than on the raw input data that the adversary seeks to recover.

To circumvent this limitation, adversaries have adopted two primary approaches. The first, more direct method modifies the network architecture by inserting malicious models near the network's input~\cite{loki, rtf}. While this allows for the direct reconstruction of raw inputs, such architectural changes are conspicuous and easily detectable by clients~\cite{garov2024hiding, wang2025hear}. A more sophisticated and stealthy approach proposed by Shi et al.~\cite{Shi2023ScaleMIAAS} avoids altering the architecture. Instead, it applies manipulations to existing linear layers within the classifier to reconstruct the LSRs. Subsequently, the attacker employs a server-side decoder to invert the feature extractor's function, mapping the recovered representations back to the original input space~\cite{Shi2023ScaleMIAAS}.

\subsection{Paired Weight GIAs}
The paired weight attack~\cite{mkor} is another sophisticated handcrafted \gls{gia} that manipulates consecutive linear \gls{fc} layers to reconstruct their input perfectly.

\smallskip
\noindent\textbf{Intuition and Attack Steps.}
The core strategy is to structure the weights and biases of a linear layer such that for any input, only one neuron within each designated pair can activate. This deterministic activation pattern allows the adversary to analytically invert the gradient computation. Consider the first linear layer with weights $W_1 \in \mathbb{R}^{k \times d}$ and bias $b_1 \in \mathbb{R}^k$, where $k$ is even.
\begin{enumerate}
    \item \textbf{Paired Weight Configuration:} The attacker pairs neurons $(n-1, n)$ for all even $n \in \{2, 4, \ldots, k\}$. The weight row for neuron $n$ is set to a negatively scaled version of the weight row for neuron $n-1$: $W_{1,n,:} = \alpha_n W_{1,n-1,:}$, where $\alpha_n < 0$.
    \item \textbf{Coordinated Bias Setting:} The biases are set similarly: $b_{1,n} = \alpha_n b_{1,n-1}$. Consequently, the pre-activations are related by $a_n = \alpha_n a_{n-1}$. Since $\alpha_n$ is negative, $a_n$ and $a_{n-1}$ have opposite signs (or are zero). After applying a ReLU activation, at most one of them can be non-zero.
    \item \textbf{Merging Path Design:} Subsequent layers are configured with structured weights (e.g., block-diagonal or positive-only matrices) to ensure that the gradient flow remains analytically tractable and invertible.
\end{enumerate}

\smallskip
\noindent\textbf{From Feature Reconstruction to Input Recovery.}
The paired weight mechanism described above enables the perfect reconstruction the manipulated layers' input, which are the intermediate LSRs in a standard \gls{cnn}. 
However, this alone is insufficient for recovering the original network input. The feature extractor is a lossy function that discards spatial information, making its output non-invertible by default. To overcome this, the paired weight attack is a two-phase strategy that also involves manipulating the feature extractor itself. As described by Wang et al.~\cite{mkor}, the adversary replaces standard convolutional filters with custom-designed pairs of filters that are engineered to preserve spatial details. This ensures that the resulting LSRs are not only perfectly reconstructible but also retain sufficient fine-grained information about pixel locations and values, allowing for a complete and analytical inversion back to the original input data.
We refer to the original paper~\cite{mkor} and Appendix~\ref{sec:appendix_paired} for a full description of this approach.

\smallskip
\noindent\textbf{Transformation Function.}
The transformation rules for the paired weight attack can be formalized as:
\begin{equation}
\mathcal{R}_{\text{paired}} = \left\{
    \begin{alignedat}{2}
        &W_{1, n, :} = \alpha_n W_{1, n-1, :} \quad  \forall n \in \{2, 4, \ldots, k\} \\
        &b_{1, n} = \alpha_n b_{1, n-1} \quad  \forall n \in \{2, 4, \ldots, k\} \\
        & W_l, b_l \text{ structured } \quad \forall l > 1
    \end{alignedat}
\right.
\end{equation}
where $\alpha_n < 0$ are negative scaling factors. The transformation is then $\theta^* = \mathcal{T}(\theta; \mathcal{R}_{\text{paired}})$.

\smallskip \noindent\textbf{Reconstruction Function.}
The deterministic activation pattern allows the attacker to solve for the input features from the gradients. For an attack targeting the final classifier, where each neuron corresponds to a class, the input feature vector $\hat{z}$ for a specific class can be recovered. A simplified reconstruction function for a batch of $m$ samples is:
\begin{equation}
    \hat{z}^T = \mathcal{F}(\nabla_{\theta^*} \mathcal{L}) = \left. \sum_{k=1}^{m} \nabla_{W_{1,y_k,:}} \mathcal{L} \right/ \sum_{k=1}^{m} \nabla_{b_{1,y_k}} \mathcal{L}
\end{equation}
where $y_k$ is the label of the $k$-th sample $x_k$, and the gradients are taken with respect to the parameters of the neuron corresponding to that label. This formula recovers a batch-averaged feature vector by exploiting the same gradient-to-input proportionality as in the binning attack.

\begin{figure}[t]
    \centering
    \begin{adjustbox}{max width=\columnwidth}
        \includegraphics[width=0.75\columnwidth]{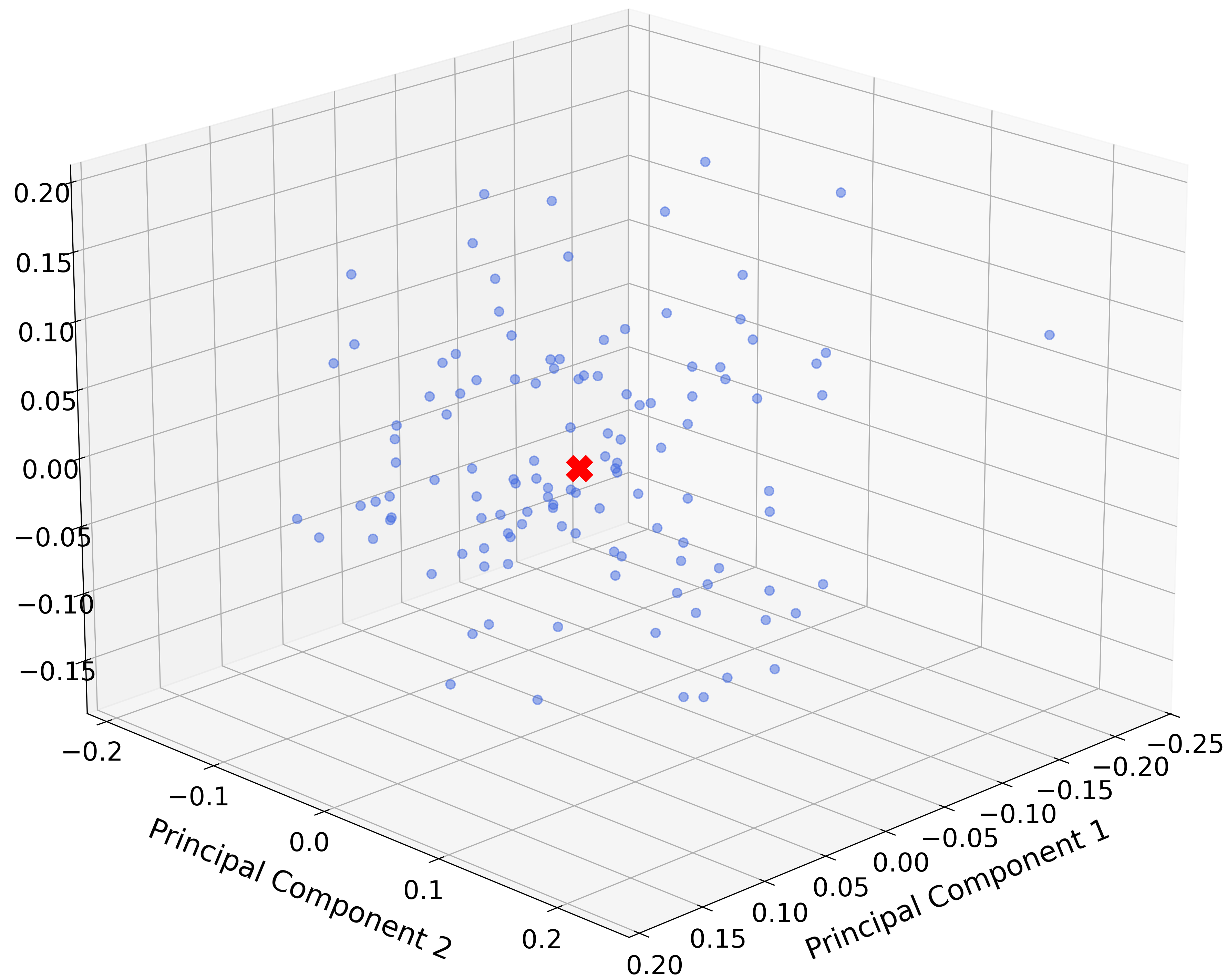}
    \end{adjustbox}
    \caption{Neuron diversity collapse visualized via 3D PCA. Benign neurons (blue) create a scattered cloud, indicating varied features. All neurons of the model manipulated with Shi et al.~\cite{Shi2023ScaleMIAAS} collapse to a single (red) point, showing the attack forces identical neuron representations.}   
    \label{fig:neuron_diversity}
\end{figure}

\subsection{Detection of Handcrafted GIAs}
Our analysis of handcrafted \glspl{gia} reveals that their underlying manipulation logic primarily rely on the static assignment of weights within the linear layers. These manipulations predominantly target the first two linear layers of the global model's classifier, previously denoted as $W_1$ and $W_2$. 

Furthermore, an examination of both the $\mathcal{R}_{binning}$ and $\mathcal{R}_{paired}$ rule sets highlights a critical point that is largely overlooked in the existing literature: \textbf{both approaches force all neurons within a layer to compute either the same function or linearly correlated functions}. 
In this way, the diversity of each neuron collapses into a single (or several) functions reaching a configuration which is highly improbable during the standard training of a \gls{ml} model, where each neuron is expected to learn a distinct and independent representation. Figure~\ref{fig:neuron_diversity} provides a high-level visualization of the neuron diversity collapse induced by Shi et al.~\cite{Shi2023ScaleMIAAS}. 

Algorithm~\ref{alg:handcrafted_detection} details the client-side detection process for handcrafted \glspl{gia}. The methodology is based on iterating through each linear layer of the global model and applying a series of statistical tests to its weight matrix $W$ and bias vector $b$ to identify anomalous, non-random patterns indicative of an attack.

Our detection logic relies on four primary anomaly scores, designed to capture the effects of handcrafted \glspl{gia}. The Neuron Diversity ($D$) score quantifies the dissimilarity among neurons by computing the average pairwise Euclidean distance between their weight vectors, $W^{(i)}$ and $W^{(j)}$. In attacks where neurons are either identical or very similar, this value approaches zero, contrasting sharply with the diverse representations learned in a legitimate model. Simultaneously, the Rank Ratio ($R$) measures the linear independence of the neuron weights. Handcrafted attacks that force neurons to compute linearly correlated functions will result in a rank-deficient weight matrix $W$, causing $R$ to be significantly less than 1. Furthermore, the Weight Entropy ($H$) assesses the complexity of the weight distribution after discretization. Manipulated models often use a small, predictable set of weights, leading to a low-entropy distribution, whereas trained weights typically exhibit higher randomness.
In addition to the weights, the Bias Anomaly ($B$) check is designed to detect structured, non-random bias vectors. This score combines two checks: one term ($B_m$) flags biases that are perfectly sorted, while the second ($B_s$) detects biases that exhibit regular, constant spacing between consecutive elements. The final decision step aggregates these findings. 

\takeaway{hand_gias}{T-1}{Takeaway on Handcrafted \glspl{gia}.}{
While avoiding architectural manipulation is a critical step toward creating stealthier handcrafted \glspl{gia}, it is insufficient for achieving attack stealthiness. Despite claims in previous works~\cite{Shi2023ScaleMIAAS,mkor}, these attacks still introduce highly structured weight configurations that are statistically improbable to emerge from a legitimate training.
}

\begin{algorithm}[t!]
\caption{Client-side Detection of Handcrafted \glspl{gia}}
\footnotesize
\label{alg:handcrafted_detection}
\begin{algorithmic}[1]
\Require Shared model $f_\theta$; anomaly thresholds $\tau_D$, $\tau_H$, $\tau_R$, $\tau_B$
\For{each linear layer $L$ in $f_\theta$}
    \State \textbf{Extract:}
    \State \quad Weight matrix $W \in \mathbb{R}^{n \times d}$
    \State \quad Bias vector $b \in \mathbb{R}^n$
    \vspace{0.5em}
    \State \textbf{Compute anomaly scores:}
    \State \quad \textit{Neuron diversity:}
    \Statex \quad \quad \quad $D = \frac{1}{n(n-1)} \sum_{i \neq j} \| W^{(i)} - W^{(j)} \|_2$ \vspace{0.3em}
    \State \quad \textit{Weight entropy:}
    \Statex \quad \quad \quad $H = -\sum_{i=1}^{B} p_i \log_2 p_i$ \vspace{0.3em}
    \State \quad \textit{Rank ratio:}
    \Statex \quad \quad \quad $R = \frac{\text{rank}(W)}{\min(n, d)}$ \vspace{0.3em}
    \State \quad \textit{Bias anomaly:}
    \Statex \quad \quad \quad Monotonic ordering:
    \Statex \quad \quad \quad \quad \[
        B_m = \mathbb{I}\left[ b^{(1)} < b^{(2)} < \ldots < b^{(n)} \right]
    \]
    \Statex \quad \quad \quad Regular spacing:
    \Statex \quad \quad \quad \quad \[
        B_s = \mathbb{I}\left[ \max_{i} \left| b^{(i+1)} - b^{(i)} - \Delta_b \right| < \tau_B \right]
    \]
    \Statex \quad \quad \quad $B = B_m + B_s$
    \vspace{0.5em}
    \State \textbf{Decision:}
    \If{$D < \tau_D$ \textbf{ or } $H < \tau_H$ \textbf{ or } $R < \tau_R$ \textbf{ or } $B \geq 1$}
        \State \textbf{Abort} local training and notify the server
        \State \textbf{Exit}
    \EndIf
\EndFor
\vspace{0.5em}
% \State Proceed with regular local training
\end{algorithmic}
\end{algorithm}

\section{Analysis of Learned GIAs} \label{sec:analysis_learned_gias}
Recently, a new class of \glspl{gia} has emerged in which the attacker does not rely on handcrafted manipulation of the global model's weights. Instead, those manipulations are \emph{learned} through an optimization process that leverages an auxiliary dataset $D_{aux}$ to automatically discover weight updates that best serve the attacker’s objective. In other words, the attack is expressed as a data-driven optimization problem that searches for the optimal perturbation to the global model's parameter:
\begin{equation}\label{eq:learned_gias}
\begin{split}
    \theta^\mathcal{A} = \arg\min_{\theta} \, &\ \mathcal{L}_{\text{attack}} \left( \theta  ; D_{aux} \right)
\end{split}
\end{equation}
Once the optimal malicious weights $\theta^\mathcal{A}$ have been computed, they are distributed to the victim client(s) during the attack round. This learned-weight strategy provides several advantages for the attacker: \textit{(i)} the learned malicious weights may not exhibit easily detectable patterns, thereby enhancing stealthiness against traditional anomaly detectors and aggregation-based defenses; and \textit{(ii)} the attack can be highly specific, meaning that the manipulated weights $\theta^\mathcal{A}$ can be tuned for target objectives or to facilitate the reconstruction of particular samples that possess properties of interest to the attacker. In this section, we analyze the attack logic proposed by Garov et al.~\cite{garov2024hiding} and Shan et al.~\cite{shan2025geminio} with a specific focus on client-side attack detectability.

\subsection{Secret Embedding and Reconstruction}

A representative instance of learned \glspl{gia} is the attack proposed by Garov et al.~\cite{garov2024hiding}, which introduces the SEER paradigm. They formulate the attack as a joint optimization problem involving three components: a malicious global model $f$, a secret disaggregator $d$, and a secret reconstructor $r$.
The attack is designed to enable selective data extraction from large client batches by exploiting the linear properties of gradient aggregation. Its core idea consists of training the malicious global model $f$ such that the resulting client gradients implicitly encode information about individual samples in selectively isolated and recoverable form. Specifically, the server defines a secret property $\mathcal{P}$ (e.g.,~\cite{garov2024hiding} uses brightness below a threshold) chosen so that typically only one sample per client batch satisfies it. During the attack round, the three components operate as follows: 
\begin{enumerate}
    \item \textbf{Encoding}: The malicious model $f$ processes local data and produces gradients that embed sample-level information according to the hidden encoding scheme.
    \item \textbf{Disaggregation}: The secret disaggregator $d$ projects the global models' aggregated gradients into a latent space, canceling out contributions from samples that do not satisfy $\mathcal{P}$ while preserving the target sample’s signal.
    \item \textbf{Reconstruction}: The secret reconstructor $r$ takes the input of the disaggregator, and tries to reconstruct the original input.
\end{enumerate}

Unlike earlier strategies that explicitly modify shared models' gradients, SEER performs all manipulations within a hidden subspace known only to the attacker. This concealment renders the attack substantially harder to detect using standard anomaly-based or statistical defenses like \gls{dsnr}. The optimal configuration $\theta^\mathcal{A}$ is obtained by solving:
\begin{equation}
\theta^\mathcal{A}_f, \theta^\mathcal{A}_d, \theta^\mathcal{A}_r = \arg\min_{\theta_f, \theta_d, \theta_r} \, \mathcal{L}_{\text{rec}} + \alpha \cdot \mathcal{L}_{\text{nul}} \label{eq:seer}
\end{equation}

where $\theta^\mathcal{A}_f, \theta^\mathcal{A}_d, \theta^\mathcal{A}_r$ denote the attacker-learned parameters of the malicious model $f$, the disaggregator $d$ and the reconstructor $r$. $\mathcal{L}_{\text{rec}}$ is a reconstruction loss that measures how well $r$ recovers samples satisfying the secret property $\mathcal{P}$, while $\mathcal{L}_{\text{nul}}$ is a nullification loss that drives gradients of non-target samples toward the null space of $d$. The parameter $\alpha > 0$ balances reconstruction fidelity against the degree of nullification.

SEER exploits \gls{bn} statistics to amplify per-sample signal while keeping manipulations implicit in a hidden subspace, and it is usually configured so that only one sample per client batch satisfies $\mathcal{P}$. In contrast to handcrafted attacks, this learned strategy aims to satisfy formal stealthiness requirements by avoiding both \textit{(i)} obvious signatures in gradient space through latent-space disaggregation, and \textit{(ii)} simplistic weight-space artifacts because the malicious behaviour is discovered via optimization rather than explicitly injected.

\subsection{Language-Guided Gradient Inversion}
A further representative example of learned \gls{gia} is Geminio, proposed by Shan et al.~\cite{shan2025geminio}, which introduces the first natural language interface for targeted \glspl{gia}. Rather than relying on predefined properties or coarse class labels, Geminio enables attackers to describe valuable data using natural language queries (e.g., "any weapons", "human faces"), greatly increasing the flexibility and expressiveness of targeted reconstruction. Geminio formulates the attack as a \gls{vlm}-guided loss surface reshaping problem. Given an attacker-specified query $\mathcal{Q^\mathcal{A}}$ and an auxiliary dataset $D_{aux}$, a pretrained \gls{vlm} measures the semantic alignment between each auxiliary sample and $\mathcal{Q^\mathcal{A}}$. 
These alignment scores are converted into per-sample weights and injected into the attacker's training objective. Consequently, the loss contribution is amplified for samples matching $\mathcal{Q^\mathcal{A}}$, while the loss for non-matching samples is suppressed. Formally, the attacker optimizes:

\begin{equation}
\theta^\mathcal{A} = \arg\min_{\theta} \, \mathcal{L}_{\text{Geminio}}(\theta; \mathcal{Q^\mathcal{A}}, D_{aux})\label{eq:geminio}
\end{equation}

where, for an auxiliary batch \(\mathcal{B}_{\text{aux}}\subset D_{\text{aux}}\),
\begin{equation}
\mathcal{L}_{\mathrm{Geminio}}(\theta;\mathcal{Q^\mathcal{A}},\mathcal{B}_{\text{aux}})
\;=\;\sum_{x\in\mathcal{B}_{\text{aux}}} w_{\mathcal{Q^\mathcal{A}}}(x)\,\ell\big(f_{\theta}(x)\big).
\end{equation}

Here \(w_{{Q^\mathcal{A}}}(x)\) is the normalized VLM alignment weight for sample \(x\) (for instance, obtained via a softmax over similarity scores) and \(\ell\) denotes the per-sample training loss. Geminio is performed without handcrafted gradient manipulations or unusual network architectures. Intuitively, this attack reshapes the model's loss landscape so that, during honest local training, gradients are dominated by contributions from samples that match the query $\mathcal{Q^\mathcal{A}}$. The malicious server can then apply standard reconstruction/optimization techniques to the aggregated gradients and recover high-quality images of the targeted samples, even from large client batches.

\subsection{Detectability of Learned GIAs}
\giu{
To the best of our knowledge, the detectability of learned \glspl{gia}~\cite{garov2024hiding,shan2025geminio} remains largely underexplored in the current literature~\cite{our_sok}, especially for the work of Shan et al.~\cite{shan2025geminio}. Existing evaluations have not systematically assessed these attacks under realistic \gls{fl} conditions. Instead, the focus has primarily centered on reconstruction performance (e.g., image fidelity) rather than on practical stealth or potential detection vectors. This motivated a more comprehensive study on client-side detectability.
}

\begin{figure}[t!]
    \centering
     \includegraphics[width=0.45\textwidth]{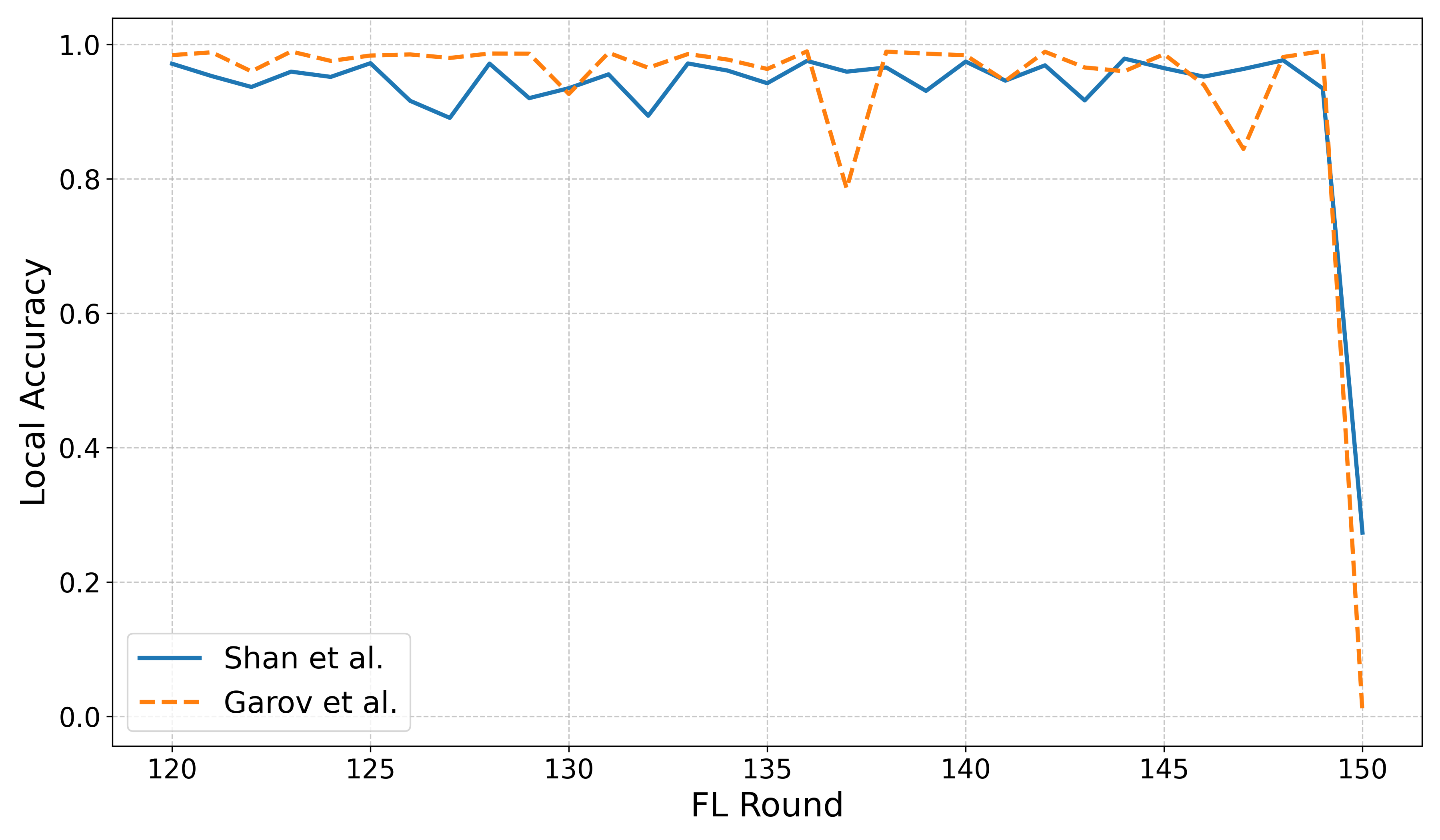}
    
    \caption{Victim client local accuracy (64 samples) showing a severe degradation at round 150, when the server distributes a manipulated model $\theta^{\mathcal{A}}_{150}$ based on Garov et al.~\cite{garov2024hiding} and Shan et al.~\cite{shan2025geminio}.}
    \label{fig:performance_drop} 
\end{figure}

During our analysis, we observed a \textit{behavioral divergence} between the manipulated global model $\theta^\mathcal{A}$ distributed by the server and the legitimate, locally-trained model $\theta_{t-1}$. Indeed, as shown in Figure~\ref{fig:performance_drop}, when the server distributes a manipulated model $\theta^\mathcal{A}$ instead of the legitimate one, the accuracy on the local client data collapse, highlighting an unexpected performance behaviour. We are the first to systematically examine this behavioral gap to develop client-side detection algorithms.

This insight guided us towards the design of a novel detection mechanism to explicitly quantify this phenomenon. Specifically, we analyze two primary dimensions during the client's local training phase: the \textbf{local loss dynamics} and the \textbf{gradient characteristics}. These indicators capture how the manipulated model deviates from the established training behavior. 

\begin{algorithm}[t!]
\caption{Loss-based Anomaly Analysis}
\label{alg:loss_analysis}
\footnotesize
\begin{algorithmic}[1]
\Require Global model $f_{\theta_t}$, previous model $f_{\theta_{t-1}}$, client data $\mathcal{D}_c$, loss $\mathcal{L}$
\Require Thresholds: $\tau_{\ell_{\text{max}}}$, $\tau_{\max\_inc}$, $\tau_{\text{spikes}}$, $\tau_{p_{95}}$, $\tau_{\text{cv\_ratio}}$, $\tau_{\text{count}}$

\State Compute loss sets $\{\ell_i^{(t)}\}$ and $\{\ell_i^{(t-1)}\}$ on $\mathcal{D}_c$.
\State For each set ($k \in \{t, t-1\}$), calculate statistics:
\Statex \quad -- Mean $\mu_\ell^{(k)}$, Std Dev $\sigma_\ell^{(k)}$, Max $\ell_{\max}^{(k)}$, 95th Percentile $p_{95}^{(k)}$
\Statex \quad -- Coefficient of Variation $\text{CV}_\ell^{(k)} = \sigma_\ell^{(k)} / \mu_\ell^{(k)}$

\Statex
\State \textbf{Compute statistical ratios:}
\Statex \quad -- Max loss increase ratio: $r_{\ell_{\max}} \gets \ell_{\max}^{(t)} / \ell_{\max}^{(t-1)}$
\Statex \quad -- P95 increase ratio: $r_{p_{95}} \gets p_{95}^{(t)} / p_{95}^{(t-1)}$
\Statex \quad -- CV increase ratio: $r_{\text{CV}} \gets \text{CV}_\ell^{(t)} / \text{CV}_\ell^{(t-1)}$
\Statex \quad -- Distributed spikes ratio: $r_{\text{spikes}} \gets |\{i: \ell_i^{(t)} > \mu_\ell^{(t-1)} + 3\sigma_\ell^{(t-1)}\}| / |\mathcal{D}_c|$

\Statex
\Statex \textit{$\triangleright$ 1. Detects an extreme loss spike}
\State $A_1 \gets \mathbb{I}\left[ \ell_{\max}^{(t)} > \tau_{\ell_{\max}} \land r_{\ell_{\max}} > \tau_{\max\_inc} \right]$

\Statex \textit{$\triangleright$ 2. Detects if spike ratio exceeds threshold}
\State $A_2 \gets \mathbb{I}\left[ r_{\text{spikes}} > \tau_{\text{spikes}} \right]$

\Statex \textit{$\triangleright$ 3. Detects significant growth in the loss tail}
\State $A_3 \gets \mathbb{I}\left[ r_{p_{95}} > \tau_{p_{95}} \right]$

\Statex \textit{$\triangleright$ 4. Detects a sharp increase in relative variability}
\State $A_4 \gets \mathbb{I}\left[ r_{\text{CV}} > \tau_{\text{cv\_ratio}} \right]$

\Statex
\If{$\sum_{j=1}^4 A_j \geq \tau_{\text{count}}$}
    \State \textbf{Flag} model as anomalous and \textbf{abort} training.
\EndIf
\end{algorithmic}
\end{algorithm}

\smallskip \noindent \textbf{Loss-based Analysis.}
\giu{
The first client-side defense mechanism is a loss-based anomaly analysis, based on the rationale that active \glspl{gia} must fundamentally alter the global model to execute the attack, thereby disrupting its learned behavior. This intuition is drawn directly from the attack logic under consideration. Geminio~\cite{shan2025geminio} explicitly manipulates the shared model $\theta^\mathcal{A}$ to amplify the loss on specific samples matching an attacker-defined query. SEER~\cite{garov2024hiding} implicitly modifies the model to produce gradients that facilitate inversion by leveraging the disaggregator and reconstructor. Despite their different goals, both strategies forcibly alter the model's established loss landscape in a manner that is detectable on the client's local data.
}

\giu{
Guided by this, we introduce the multi-faceted statistical analysis detailed in Algorithm~\ref{alg:loss_analysis}. The method compares the per-sample loss distribution of the newly received model $\theta_t$ against that of the previous, trusted model $\theta_{t-1}$. Our framework is designed to capture the behavioral signatures of an attack using four complementary checks. It detects an extreme spike in the maximum loss value ($A_1$), which serves as a direct detector for query-based attacks like Geminio that are explicitly designed to create such high-magnitude outliers. This check is complemented by monitoring for a systemic increase in the population of outlier samples ($A_2$), which identifies if the manipulation has pushed a significant portion of the client's data into an anomalous high-loss region. Furthermore, the analysis quantifies the magnitude of this distortion by tracking significant growth in the tail of the loss distribution via the 95th percentile ($A_3$). Finally, it detects a sharp increase in the relative variability of the losses ($A_4$), measured by the Coefficient of Variation ($\text{CV}_\ell^{(k)}$), signaling that the model's behavior has become erratic and inconsistent with legitimate training.
}

\begin{figure}[t!]
    \centering
    
    \includegraphics[width=0.35\textwidth]{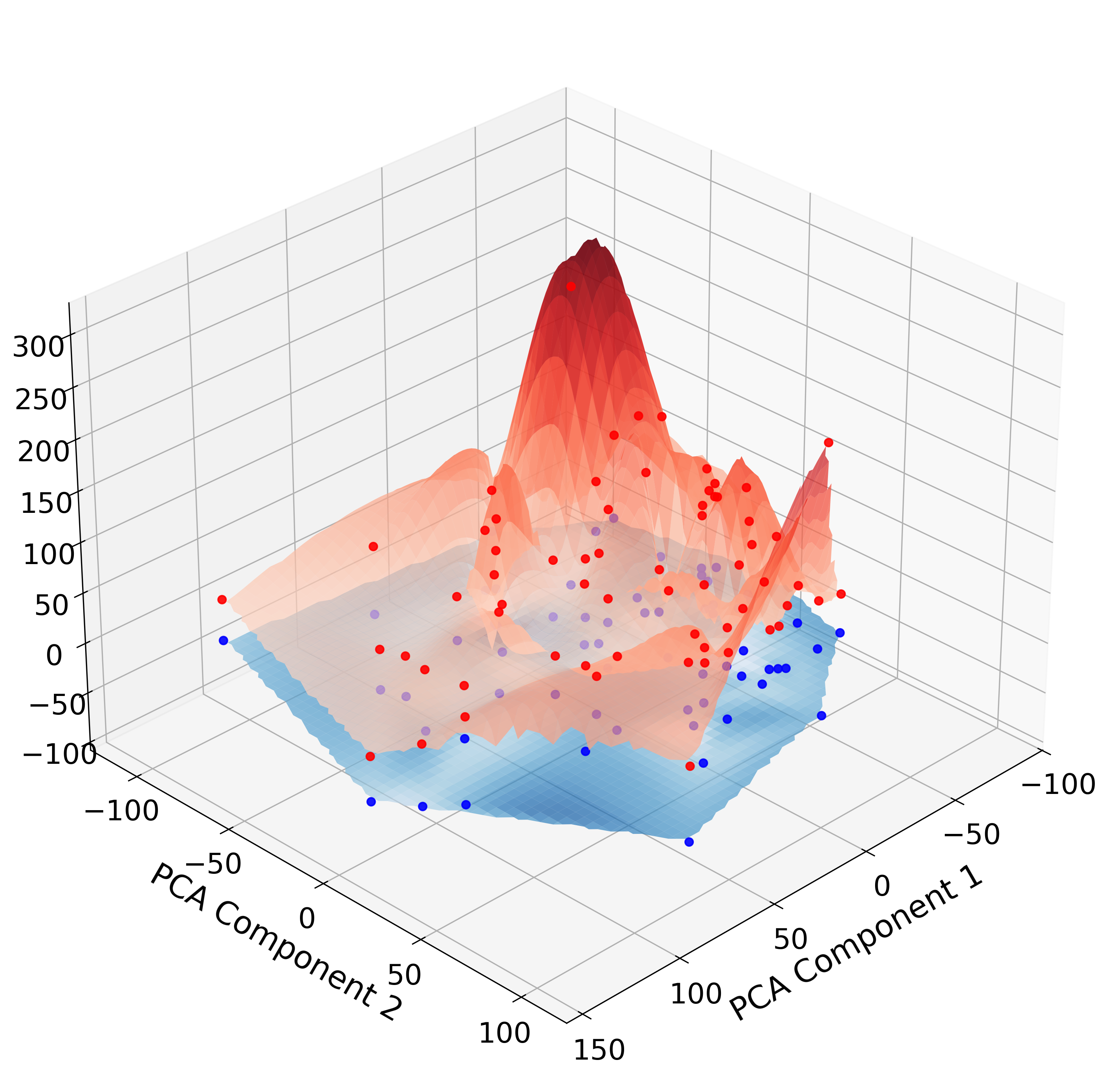}
    
    \caption{3D comparison of loss surfaces. The loss of a legitimate model is the \textbf{\textcolor{blue}{blue pre-manipulation surface}} (e.g., the legitimate model $\theta_{t-1}$), contrasted with the outcomes of a \textbf{\textcolor{figRed}{manipulated model}} with Shan et al.~\cite{shan2025geminio} (e.g., when $\theta_t$ is obtained with Eqn.~\ref{eq:geminio}).}
    \label{fig:loss_surface} 
\end{figure}

\giu{
To enhance robustness and minimize false positives, a model is flagged as anomalous and training is aborted only if at least $\tau_{\text{count}}$ (e.g., two) of these conditions are simultaneously satisfied. Figure~\ref{fig:loss_surface} provides empirical validation for this approach by illustrating the impact of the Geminio~\cite{shan2025geminio} attack on the surface loss of the shared model. A stark contrast becomes evident. The legitimate model yields a relatively smooth, low-magnitude loss landscape. In contrast, the manipulated model produces a volatile landscape characterized by sharp, high-magnitude spikes. This pronounced disruption in the loss topology is exactly what the statistical checks in Algorithm~\ref{alg:loss_analysis} are designed to identify and quantify.
}

\smallskip \noindent \textbf{Gradient-based Analysis.}
\giu{
Our second client-side check directly monitors gradient characteristics. This is motivated by the fact that both attacks, despite with different goals, must manipulate the input-to-gradient mapping. For example, SEER~\cite{garov2024hiding} relies on a server-side disaggregator and reconstructor that are \textit{jointly trained} with the shared model $\theta^\mathcal{A}$. This co-adaptation process inherently constrains $\theta^\mathcal{A}$ to produce gradients with a structure that is easy to invert for the disaggregator and reconstructor. While Geminio~\cite{shan2025geminio} also manipulates gradients to encode information, the gradient analysis we propose is specifically designed to detect the unique signature of SEER: a pronounced \textit{suppression} of the gradient signal.
}

\giu{
Our proposed divergence-based approach is essential for detecting sophisticated attacks like SEER~\cite{garov2024hiding}. Previous metrics, such as \gls{dsnr}~\cite{garov2024hiding}, are ineffective because they perform a static analysis, examining the gradient of the manipulated model $\theta^\mathcal{A}$ in isolation. SEER is explicitly engineered to defeat such checks, as it leverages a secret, server-side decoder to avoid disaggregation in the gradient space~\cite{garov2024hiding}. This ensures the resulting gradients appear benign, making their \gls{dsnr} values indistinguishable from those of legitimate networks~\cite{garov2024hiding}. We posit that this gradient-space stealth is achieved at the cost of a significant and detectable \textit{behavioral divergence}. By shifting the analysis from a static gradient snapshot to the temporal divergence between the current malicious model $\theta_t^\mathcal{A}$ and the previous trusted model $\theta_{t-1}$, our methodology bypasses the fundamental limitations of prior detection techniques.
}

\begin{figure}[t!]
    \centering

    \includegraphics[width=0.40\textwidth]{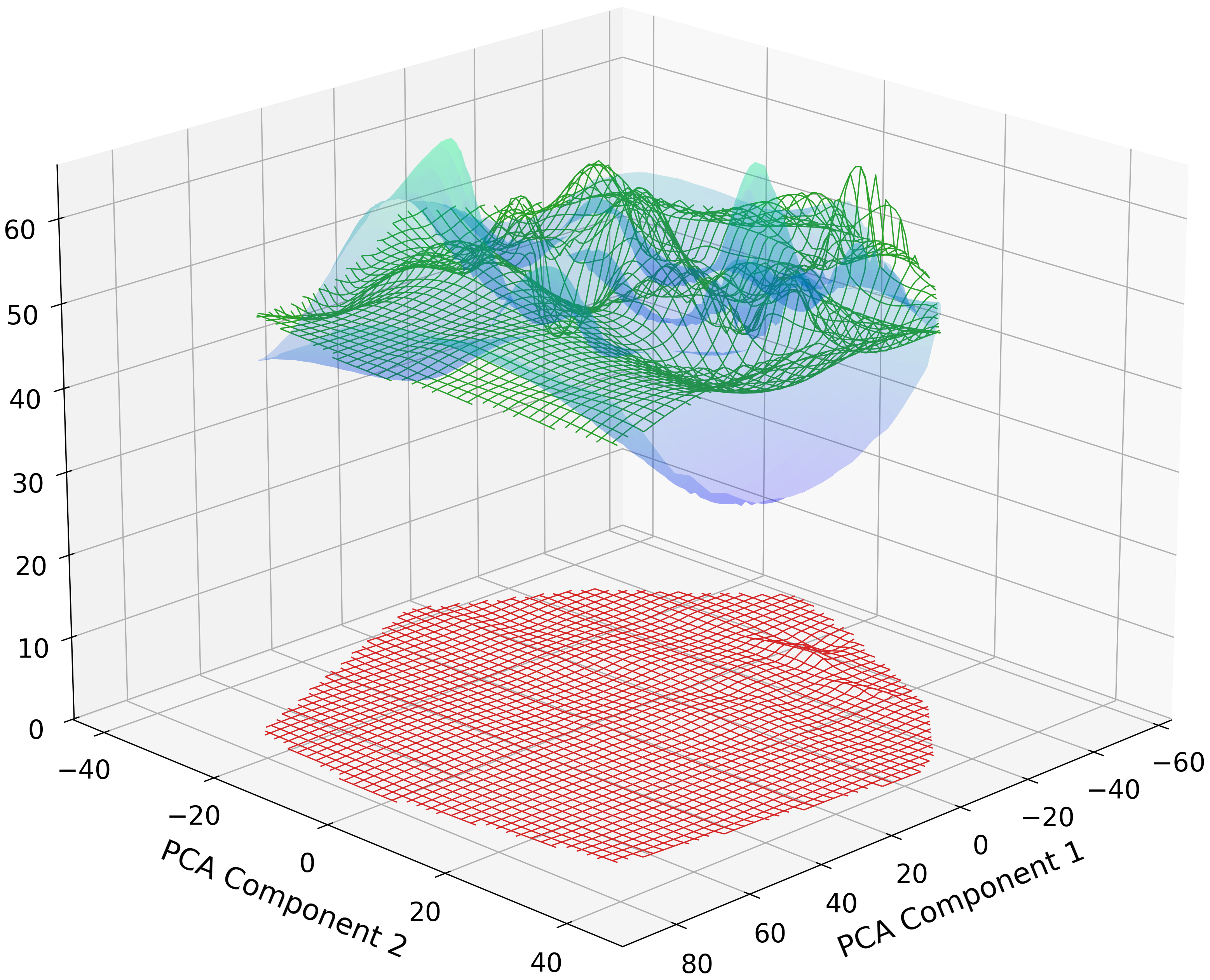}
    
    \caption{Comparison of $L_2$ gradient norm surfaces. The plot contrasts the initial legitimate model ($\theta_{t-1}$), shown as a solid \textbf{\textcolor{cyan}{pre-manipulation}} surface, with the wireframe surfaces representing a \textbf{\textcolor{figGreen}{legitimate}} update ($\theta_t$) and a \textbf{\textcolor{figRed}{manipulated}} update ($\theta^{\mathcal{A}}_t$ from Eqn.~\ref{eq:seer}).}
    \label{fig:grad_surface} 
\end{figure}

\giu{
To detect this behavior, we introduce the client-side check detailed in Algorithm~\ref{alg:gradient_analysis}. The mechanism analyzes the distribution of the $L_2$ norms of per-sample gradients, comparing the statistics from the new model $f_{\theta_t}$ against those from $f_{\theta_{t-1}}$. Our approach is tailored to identify the anomalous suppression of the gradient signal characteristic of SEER. To ensure robust detection, a model is flagged if it satisfies at least $\tau_{\text{count}}$ of the following conditions: \textit{(i)} a significant relative drop in the mean gradient norm ($B_1$), \textit{(ii)} a sharp decrease in the variability of gradient norms ($B_2$), and \textit{(iii)} a near-total collapse of the average gradient norm ($B_3$).
}
\giu{
The effect of this manipulation is visually confirmed in Figure~\ref{fig:grad_surface}. This figure contrasts the 3D gradient norm landscapes, revealing the manipulated model's surface to be conspicuously flat. This demonstrates a massive suppression of the gradient signal when compared to the previous legitimate model and the expected legitimate model (which is unknown to the client but exhibits a similar, high-magnitude landscape).
}

\takeaway{learned_gias}{T-2}{Takeaway on Learned \glspl{gia}.}{
While learned \glspl{gia} mark a significant step forward compared to handcrafted techniques, existing approaches~\cite{garov2024hiding, shan2025geminio} still fall short of evading client detectability. This limitation persists even when attacks are explicitly designed to satisfy formal stealth guarantees related to the gradient space of the manipulated model~\cite{garov2024hiding}. The key oversight lies in neglecting behavioral side-channels, as clients can still detect the presence of an attack by observing statistically significant and improbable deviations in the model’s local behavior, an aspect that current methods fail to conceal.
}

\begin{algorithm}[t!]
\caption{Gradient-based Analysis}
\label{alg:gradient_analysis}
\footnotesize
\begin{algorithmic}[1]
\Require Global model $f_{\theta_t}$, previous model $f_{\theta_{t-1}}$, client data $\mathcal{D}_c$, loss $\mathcal{L}$
\Require Thresholds: $\tau_{g_{\text{norm}}}$, $\tau_{g_{\text{var}}}$

\State Compute per-sample gradients $\{g_i^{(t)}\}$ and $\{g_i^{(t-1)}\}$ on the entire client dataset $\mathcal{D}_c$.

\State Calculate statistics based on the L2 norms of the gradients:
\Statex \quad -- Mean of norms: $\mu_g^{(k)}$ for $k \in \{t, t-1\}$
\Statex \quad -- Standard deviation of norms: $\sigma_g^{(k)}$ for $k \in \{t, t-1\}$
\Statex \quad -- Relative reduction in mean:
\Statex \quad \quad \quad $r_{\text{norm}} = (\mu_g^{(t-1)} - \mu_g^{(t)}) / \mu_g^{(t-1)}$
\Statex \quad -- Relative reduction in std dev:
\Statex \quad \quad \quad $r_{\text{var}} = (\sigma_g^{(t-1)} - \sigma_g^{(t)}) / \sigma_g^{(t-1)}$

\Statex
\Statex \textit{$\triangleright$ Checks for an anomalous drop in the average gradient norm}
\State $B_1 \gets \mathbb{I}[r_{\text{norm}} > \tau_{g_{\text{norm}}}]$

\Statex \textit{$\triangleright$ Checks for an anomalous drop in gradient norm variability}
\State $B_2 \gets \mathbb{I}[r_{\text{var}} > \tau_{g_{\text{var}}}]$

\Statex \textit{$\triangleright$ Checks for a collapse in the average gradient norm}
\State $B_3 \gets \mathbb{I}[\mu_g^{(t)} < 0.1 \cdot \mu_g^{(t-1)}]$

\Statex
\If{$B_1 + B_2 + B_3 \geq 2$}
    \State \textbf{Flag} model as anomalous and \textbf{abort} training.
\EndIf
\end{algorithmic}
\end{algorithm}

\section{Experimental Design}\label{sec:experimental_design}
This section details the experimental framework designed to validate our client-side detection mechanisms. Unlike prior work on \glspl{gia}, which focuses on reconstruction fidelity, our evaluation emphasizes attack viability and stealthiness in realistic \gls{fl} settings. All experiments were performed on a system equipped with an NVIDIA A100-SXM4-80GB GPU.

\smallskip \noindent \textbf{Federated Setup.}
We simulate a federated network with one malicious active server and two network configurations composed of 10 and 100 clients, using the Flower framework~\cite{beutel2020flower}. To reflect a realistic vulnerability scenario, we test our algorithms when 20\% of the clients are randomly designated as victims and assigned smaller local datasets, aligning with prior findings that \glspl{gia} are more effective when gradients are computed over smaller  batches~\cite{du2024sok,our_sok}. For the sake of fairness, we replicate the original threat model of each attack in terms of victim dataset size~\cite{mkor,Shi2023ScaleMIAAS,garov2024hiding,shan2025geminio}. Furthermore, we average outcomes over simulations with different victim clients to ensure the robustness of the reported results across varying client data heterogeneity. 
\giu{Finally, we evaluate our detection algorithms under partial participation, selecting 30\% of clients per round. This setting may pose a significant challenge for divergence-based detectors, as an inactive client comparing the newly received model against their outdated local version may register a large, legitimate divergence, potentially leading to a high rate of false positives.}

\smallskip \noindent \textbf{Datasets and Data Partitioning.}
In each setup, the training set is split among clients, while the server retains the full test set for evaluating the global model's performance and as auxiliary dataset $D_{\text{aux}}$ for attacks where necessary as in Shi et al.~\cite{Shi2023ScaleMIAAS}, Shan et al.~\cite{shan2025geminio} and Garov et al.~\cite{garov2024hiding}. 
\giu{
In this section, we present the experimental results under the \gls{niid} data distribution, a common and realistic assumption in \gls{fl}.
}
For the \gls{niid} setting, we introduce data heterogeneity through distribution-based label skew via \gls{lda}~\cite{hsu2019measuring} with a concentration parameter of $\alpha=0.3$.
The datasets used in this work are CIFAR-10~\cite{krizhevsky2009learning}, CIFAR-100~\cite{krizhevsky2009learning}, TinyImageNet~\cite{deng2009imagenet}, MNIST~\cite{mnist}, FMNIST~\cite{fmnist}. For the sake of completeness, a corresponding evaluation under the \gls{iid} data distribution is provided in Appendix~\ref{sec:appendix_experimental}.

\smallskip \noindent \textbf{Model and Training Implementation.} 
We simulate an \gls{fl} training with 10 or 100 clients over 200 and 100 rounds, respectively. To ensure a fair comparison, we adopt the same neural network architectures and learning algorithms (\gls{fedavg} or \gls{fedsgd}) used in the original work proposing each attack. This allows us to assess the detectability of each attack within its intended model configuration. The architectures tested in our experiments include ResNet18~\cite{he2016deep} and ResNet34~\cite{he2016deep}, LeNet5~\cite{lecun2002gradient}, VGG16~\cite{simonyan2014very}, and a small \gls{cnn}~\cite{Shi2023ScaleMIAAS} with four convolutional layers followed by a three-layer fully connected classifier.

\begin{table}[t!]
\centering
\begin{tabular}{l r r r}
\toprule
\textbf{Parameter} & \textbf{Conservative} & \textbf{Standard} & \textbf{Aggressive} \\
\midrule

\quad $\tau_D$ & $10^{-4}$ & $10^{-3}$ & $10^{-2}$ \\
\quad $\tau_H$ & 2.0 & 3.0 & 4.0 \\
\quad $\tau_R$ & 0.5 & 0.8 & 0.9 \\
\quad $\tau_{\text{count}}$ & 2 & 2 & 2 \\
\midrule

\quad $\tau_{\ell_{\text{max}}}$ & 25.0 & 10.0 & 4.0 \\
\quad $\tau_{mi}$ & 20.0 & 10.0 & 1.0 \\
\quad $\tau_{\text{spikes}}$ & 0.5 & $10^{-1}$ & $10^{-2}$ \\
\quad $\tau_{p_{95}}$ & 3.0 & 3.0 & 0.8 \\
\quad $\tau_{cv}$ & 2.0 & 1.5 & 1.1 \\
\midrule

\quad $\tau_{g_{\text{norm}}}$ & 0.8 & 0.5 & $10^{-5}$ \\
\quad $\tau_{g_{\text{var}}}$ & 0.4 & 0.2 & $10^{-5}$ \\
\bottomrule
\end{tabular}
\caption{Comparison of Conservative, Standard, and Aggressive configurations for anomaly detection thresholds. Parameters correspond to Algorithms~\ref{alg:handcrafted_detection},~\ref{alg:loss_analysis}, and~\ref{alg:gradient_analysis}.}
\label{tab:config_parameters}
\end{table}

\smallskip \noindent \textbf{Attack Implementation.} 
We simulate an active malicious server that operates honestly except during designated \emph{attack rounds}, where the server selectively sends the manipulated model to the two designated victim clients while others receive the legitimate version. To test the detection robustness of our algorithms, we vary the attack timing across different stages of training: during the initial phase, midway through training, and after the model has converged. \giu{Although our primary experiments exclude an attack in the very first round (as initial weights in some \gls{fl} setups can be client-sourced) we also discuss the detectability of such a scenario for both classes of attacks under consideration in Appendix~\ref{sec:appendix_experimental}.}
Malicious models are crafted using the test set as the auxiliary dataset $D_{\text{aux}}$, which is used to derive the necessary statistics for the parameter manipulation. To ensure a faithful replication of each attack, we utilize the official, author-provided source code\footnote{\url{https://github.com/wfwf10/MKOR}}\textsuperscript{,}\footnote{\url{https://github.com/unknown123489/Scale-MIA}}\textsuperscript{,}\footnote{\url{https://github.com/insait-institute/SEER}}\textsuperscript{,}\footnote{\url{https://github.com/HKU-TASR/Geminio}} and adopt the same hyperparameter configurations proposed in the original works.

\smallskip \noindent \textbf{Metrics.} 
We evaluate the efficacy of the proposed detection algorithm using the \gls{tpr} and the \gls{fpr}.
In this experimental context, a positive sample is defined as a communication round wherein a client receives a manipulated model (i.e., an attack round).
Conversely, a negative sample denotes a round involving a legitimate, benign model. The \gls{tpr}, also known as sensitivity, quantifies the fraction of attacks that are correctly identified.
The \gls{fpr}, or false alarm rate, measures the fraction of benign rounds erroneously classified as attacks.
These metrics are formally defined as:
\begin{equation}
\label{eq:tpr_fpr}
\text{TPR} = \frac{\text{TP}}{\text{TP} + \text{FN}} \qquad \text{FPR} = \frac{\text{FP}}{\text{FP} + \text{TN}}
\end{equation}
where TP, FN, FP, and TN represent the total number of True Positives, False Negatives, False Positives, and True Negatives, respectively. An ideal detection system maximizes the \gls{tpr} while minimizing the \gls{fpr}.

\section{Evaluation}\label{sec:evaluation}
This section presents the evaluation of our proposed client-side detection algorithms. Our analysis aims to address the following research questions:
\begin{enumerate}
    \item What is the performance of the joint detection method (combining the checks introduced in Algorithms~\ref{alg:handcrafted_detection}, ~\ref{alg:loss_analysis}, and~\ref{alg:gradient_analysis}) in the considered realistic setup, measured by \gls{tpr} and \gls{fpr}?
    \item How robust is the performance of the detection methods to variations in their sensitivity parameters?
    \item How do the detection algorithms perform in scenarios with partial client participation?
    \item What is the client-side computational overhead introduced by the detection algorithms?
\end{enumerate}

\smallskip \noindent \textbf{Detectability.} 
Table~\ref{tab:experiment_results} presents the experimental results for our detection mechanisms across the various attack configurations, datasets, models, and \gls{fl} setups considered. In all configurations, all client-side detection algorithms (Algorithm~\ref{alg:handcrafted_detection}, Algorithm~\ref{alg:loss_analysis}, and Algorithm~\ref{alg:loss_analysis}) were enabled using the standard parameters shown in Table~\ref{tab:config_parameters}. The results demonstrate that our algorithms effectively detect attacks in both \gls{fl} configurations and across diverse attacks, datasets, and victim clients. As these values are averaged over 10 distinct runs for each configuration, we conclude that our detection algorithms are robust to variations in client data distribution and model initialization.

\begin{table*}[h!]
\centering
\small
\renewcommand{\arraystretch}{1.5} % Aumenta spaziatura righe
\begin{adjustbox}{max width=\textwidth}
% Setup(l), Cat(l), Atk(l), Data(l), Model(l), Rnds(c), BS(c), Runs(c), TPR(c), FPR(c)
\begin{tabular}{l l l l l c c c c c}
\toprule
  \shortstack{\textbf{FL Setup}\\\textbf{ }} & \shortstack{\textbf{Category}\\\textbf{ }} & \shortstack{\textbf{Attack}\\\textbf{ }} & \shortstack{\textbf{Dataset}\\\textbf{ }} & \shortstack{\textbf{Global}\\\textbf{Model}} & \shortstack{\textbf{Total}\\\textbf{Rounds}} & \shortstack{\textbf{Victim}\\\textbf{Batch Size}} & \shortstack{\textbf{Number of}\\\textbf{Runs}} & \shortstack{\textbf{Average}\\\textbf{TPR$\uparrow$}} & \shortstack{\textbf{Average}\\\textbf{FPR$\downarrow$}}  \\
\midrule
  \multirow{12}{*}{\shortstack{Cross-Silo\\(10 clients)}} &   \multirow{5}{*}{\shortstack{Handcrafted\\GIAs}} & \cite{Shi2023ScaleMIAAS} & CIFAR-10~\cite{krizhevsky2009learning} & ResNet18~\cite{he2016deep} & 200 & 64 & 10 & $1.000_{\pm0.000}$ & $0.339_{\pm0.165}$ \\
   &    & \cite{Shi2023ScaleMIAAS} & FMNIST~\cite{fmnist} & CNN~\cite{Shi2023ScaleMIAAS} & 200 & 64 & 10 & $1.000_{\pm0.000}$ & $0.243_{\pm0.218}$ \\
   &    & \cite{Shi2023ScaleMIAAS} & Tiny-ImageNet~\cite{deng2009imagenet} & CNN~\cite{Shi2023ScaleMIAAS} & 200 & 64 & 10 & $1.000_{\pm0.000}$ & $0.000_{\pm0.000}$ \\
  
  \cline{3-10}
   
   &    & \cite{mkor} & FMNIST~\cite{fmnist} & LeNet5~\cite{lecun2002gradient} & 200 & 100 & 10 & $1.000_{\pm0.000}$ & $0.000_{\pm0.000}$ \\
   &    & \cite{mkor} & MNIST~\cite{mnist} & LeNet5~\cite{lecun2002gradient} & 200 & 100 & 10 & $1.000_{\pm0.000}$ & $0.001_{\pm0.002}$ \\

  \cmidrule(l){2-10}
   &   \multirow{6}{*}{\shortstack{Learned\\GIAs}} & \cite{garov2024hiding} & CIFAR-100~\cite{krizhevsky2009learning} & ResNet18~\cite{he2016deep} & 200 & 64 & 10 & $1.000_{\pm0.000}$ & $0.000_{\pm0.000}$ \\
   &    & \cite{garov2024hiding} & CIFAR-10~\cite{krizhevsky2009learning} & ResNet18~\cite{he2016deep} & 200 & 64 & 10 & $1.000_{\pm0.000}$ & $0.073_{\pm0.081}$ \\
   &    & \cite{garov2024hiding} & Tiny-ImageNet~\cite{deng2009imagenet} & ResNet18~\cite{he2016deep} & 200 & 64 & 10 & $1.000_{\pm0.000}$ & $0.000_{\pm0.000}$ \\
  
  \cline{3-10}
  
   &    & \cite{shan2025geminio} & CIFAR-100~\cite{krizhevsky2009learning} & ResNet34~\cite{he2016deep} & 200 & 64 & 10 & $1.000_{\pm0.000}$ & $0.000_{\pm0.000}$ \\
   &    & \cite{shan2025geminio} & CIFAR-10~\cite{krizhevsky2009learning} & ResNet34~\cite{he2016deep} & 200 & 64 & 10 & $1.000_{\pm0.000}$ & $0.000_{\pm0.000}$ \\
   &    & \cite{shan2025geminio} & Tiny-ImageNet~\cite{deng2009imagenet} & ResNet34~\cite{he2016deep} & 200 & 64 & 10 & $1.000_{\pm0.000}$ & $0.000_{\pm0.000}$ \\
   
\midrule

  \multirow{11}{*}{\shortstack{Cross-Device\\(100 clients)}} &   \multirow{4}{*}{\shortstack{Handcrafted\\GIAs}} & \cite{Shi2023ScaleMIAAS} & CIFAR-10~\cite{krizhevsky2009learning} & ResNet18~\cite{he2016deep} & 100 & 64 & 10 & $1.000_{\pm0.000}$ & $0.001_{\pm0.002}$ \\
   &    & \cite{Shi2023ScaleMIAAS} & FMNIST~\cite{fmnist} & CNN~\cite{Shi2023ScaleMIAAS} & 100 & 64 & 10 & $1.000_{\pm0.000}$ & $0.001_{\pm0.001}$ \\
   &    & \cite{Shi2023ScaleMIAAS} & Tiny-ImageNet~\cite{deng2009imagenet} & CNN~\cite{Shi2023ScaleMIAAS} & 100 & 64 & 10 & $1.000_{\pm0.000}$ & $0.000_{\pm0.000}$ \\
  
  \cline{3-10}
  
   &    & \cite{mkor} & FMNIST~\cite{mnist} & LeNet5~\cite{lecun2002gradient} & 100 & 100 & 10 & $1.000_{\pm 0.000}$ & $0.000_{\pm 0.000}$ \\
   &    & \cite{mkor} & MNIST~\cite{mnist} & LeNet5~\cite{lecun2002gradient} & 100 & 100 & 10 & $1.000_{\pm0.000}$ & $0.002_{\pm0.001}$ \\
  \cmidrule(l){2-10}
   &   \multirow{6}{*}{\shortstack{Learned\\GIAs}} & \cite{garov2024hiding} & CIFAR-100~\cite{krizhevsky2009learning} & ResNet18~\cite{he2016deep} & 100 & 64 & 10 & $1.000_{\pm0.000}$ & $0.001_{\pm0.001}$ \\
   &    & \cite{garov2024hiding} & CIFAR-10~\cite{krizhevsky2009learning} & ResNet18~\cite{he2016deep} & 100 & 64 & 10 & $1.000_{\pm0.000}$ & $0.009_{\pm0.017}$ \\
   &    & \cite{garov2024hiding} & Tiny-ImageNet~\cite{deng2009imagenet} & ResNet18~\cite{he2016deep} & 100 & 64 & 10 & $1.000_{\pm0.000}$ & $0.001_{\pm0.003}$ \\
  
  \cline{3-10}
  
   &    & \cite{shan2025geminio} & CIFAR-100~\cite{krizhevsky2009learning} & ResNet34~\cite{he2016deep} & 100 & 64 & 10 & $1.000_{\pm0.000}$ & $0.000_{\pm0.000}$ \\
   &    & \cite{shan2025geminio} & CIFAR-10~\cite{krizhevsky2009learning} & ResNet34~\cite{he2016deep} & 100 & 64 & 10 & $1.000_{\pm0.000}$ & $0.000_{\pm0.000}$ \\
   &    & \cite{shan2025geminio} & Tiny-ImageNet~\cite{deng2009imagenet} & ResNet34~\cite{he2016deep} & 100 & 64 & 10 & $1.000_{\pm0.000}$ & $0.000_{\pm0.000}$ \\
\bottomrule
\end{tabular}
\end{adjustbox}
\renewcommand{\arraystretch}{1}
\caption{Summary of analyzed experiment configurations and detection results. 
All experiments simulate a non-IID data distribution (distribution-based label skew via LDA~\cite{hsu2019measuring}, $\alpha=0.3$). 
TPR/FPR values are averaged across several runs and presented as $\texttt{mean}_{\pm \texttt{std}}$. In each run, 20\% of clients are randomly selected as victims. 
Detection performance was evaluated at rounds 2, 40, 80, and 100 for the cross-silo setup, and at rounds 2, 50, 150, and 200 for the cross-device setup. The hyperparameter of the detection algorithms refers to the "Standard" configuration in Table~\ref{tab:config_parameters}.}
\label{tab:experiment_results}
\end{table*}

\begin{figure}[t]
    \centering
    \begin{adjustbox}{max width=\columnwidth}
        \includegraphics[width=0.8\columnwidth]{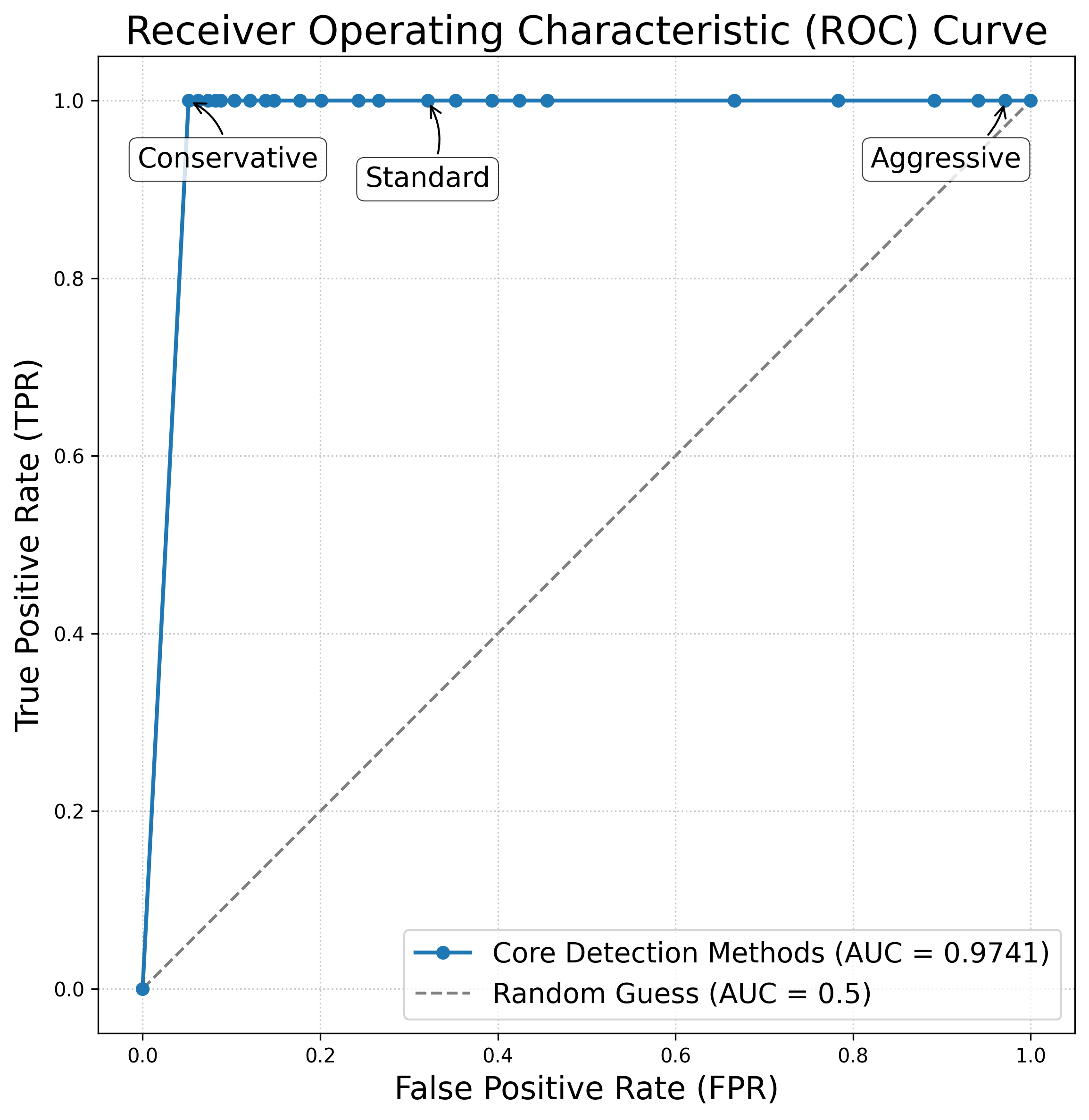}
    \end{adjustbox}
    \caption{
    ROC curve evaluating the detector's performance for the Cross-Silo setup with ResNet18 under Shi et al.~\cite{Shi2023ScaleMIAAS} attack. 
    }  
    \label{fig:roc_curve}
\end{figure} 

\giu{
Notably, our detection methods remain effective even when attacks are customized to reconstruct data satisfying specific properties. We explicitly tested this by evaluating multiple configurations of the same attack. For Garov et al.~\cite{garov2024hiding}, our results are averaged across three distinct target properties used to build the manipulated model. Similarly, for Shan et al.~\cite{shan2025geminio}, we simulated attacks by issuing queries for each class in the dataset, using the specific queries reported in the original paper~\cite{shan2025geminio} for Tiny-ImageNet. The averaged performance confirms that our detection mechanisms are robust across these varied attack instances.
}

Regarding \gls{fpr} values, our algorithms achieve very low rates across all configurations. The highest values (0.339 and 0.243) occurred in specific 10-client setups, with these false flags being mainly related to the loss analysis algorithm. However, this is highly dependent on the parameter configuration. By applying the conservative strategies from Table~\ref{tab:config_parameters}, the \gls{fpr} drops to 0.041 and 0.037, respectively, while the \gls{tpr} remains consistently near 1.0, as shown in Figure~\ref{fig:roc_curve}. This indicates that the algorithms remain capable of capturing the highly anomalous behavior of attacks, even when configured with more conservative (and then less sensitive) parameters to minimize false positives. 

\takeaway{learned_gias}{T-3}{Detectability of Active \glspl{gia}.}{
Our detection methods can successfully identify the learned \glspl{gia}~\cite{shan2025geminio,garov2024hiding} across various configurations by capturing core effects intrinsic to the attack logic, rather than relying on artifacts specific to any particular setup. The perfect \gls{tpr} values, achieved even under conservative settings, underscore the efficacy of our detection algorithms. These findings demonstrate that, despite the claimed stealthiness, their underlying attack mechanisms introduce side effects that our approach can reliably detect.
}

\smallskip \noindent \textbf{Sensitivity Analysis.} 
We conducted a sensitivity analysis on the parameters used to tune the detector's sensitivity. For brevity, we report the analysis for the experimental setup that yielded the highest \gls{tpr} values, as this represents the most effective detection scenario; however, similar sensitivity considerations hold for the other configurations evaluated in our work. To evaluate the detector's performance trade-offs, we generated a Receiver Operating Characteristic (ROC) curve by testing 30 distinct parameter configurations. These configurations were obtained by linearly interpolating the thresholds starting from those shown in Table~\ref{tab:config_parameters}. This sweep ranges from the "Conservative" setup (the permissive lower bound) to the "Aggressive" setup (the strict upper bound), using the "Standard" configuration as a baseline. The results, averaged over 10 simulation runs, demonstrate a robust classification performance, achieving an Area Under the Curve (AUC) of 0.9741. 
\giu{
Furthermore, we observe that the \gls{tpr} remains consistently near 1.0, even in the conservative setup. This high detection rate stems from our methods' ability to effectively capture the attack logic, coupled with the highly improbable configurations of the evaluated attacks.
}

\smallskip \noindent \textbf{Partial Client Participation.} 
To further validate our approach in a realistic cross-device setting, we investigated the impact of partial client participation. This scenario is a critical stress test for our model divergence analysis, as intermittent selection is common. In particular, clients that are not selected for several rounds must compare the newly received global model against a local version from a much earlier communication round. This benign divergence, stemming merely from missed updates rather than malicious manipulation, could potentially be misidentified as an anomaly, thereby inflating the \gls{fpr}. 
We also managed the edge case where a client is selected for its \textit{first time} during an attack round. In this scenario, the client compares the received (manipulated) model against a randomly initialized baseline; this mechanism also inherently covers attacks initiated by the server in the very first \gls{fl} round.

The results, summarized in Table~\ref{tab:auc_results}, demonstrate the robustness of our detection methods. In these experiments, we simulated a 30\% client participation rate per round, ensuring the victim clients were selected during the attack rounds. The resulting average values of AUC, averaged across 10 distinct runs and 30 different sensitivity parameter configurations, confirm that our approach remains highly effective. It successfully distinguishes malicious manipulations from the expected, larger divergence caused by intermittent client selection, confirming its applicability for practical cross-device deployments.

\smallskip \noindent \textbf{Client-Side Overhead.} 
Many defensive mechanisms enhance privacy at the expense of additional overhead, which, in our context, manifests as client-side computational overhead. We evaluate this overhead by performing both a theoretical analysis of computational complexity and an empirical evaluation, reporting the additional time required by the client in our experimental setup in Table~\ref{tab:overhead} of Appendix~\ref{sec:appendix_experimental}. 

\begin{table}[t!]
\centering
\small
\renewcommand{\arraystretch}{1.5} % Aumenta spaziatura righe
% Setup(l), Atk(l), Model(l), Rnds(c), AUC(c)
\begin{tabular}{l l l c c}
\toprule
  \shortstack{\textbf{FL Setup}\\\textbf{ }} & \shortstack{\textbf{Attack}\\\textbf{ }} & \shortstack{\textbf{Global}\\\textbf{Model}} & \shortstack{\textbf{Total}\\\textbf{Rounds}} & \shortstack{\textbf{Average}\\\textbf{AUC$\uparrow$}} \\
\midrule
  \multirow{2}{*}{\shortstack{Cross-Silo\\(10 clients)}} & \cite{garov2024hiding} & ResNet18 & 200 & 0.989 \\
   & \cite{shan2025geminio} & ResNet34 & 200 & 0.991 \\
   
\midrule

  \multirow{2}{*}{\shortstack{Cross-Device\\(100 clients)}} & \cite{garov2024hiding} & ResNet18 & 100 & 0.999 \\
   & \cite{shan2025geminio} & ResNet34 & 100 & 0.999 \\
\bottomrule
\end{tabular}
\renewcommand{\arraystretch}{1}
\caption{Summary of detection results (AUC) for learned \glspl{gia} on Tiny-ImageNet dataset in both cross-silo and cross-device scenarios. The configurations are the ones reported in Table~\ref{tab:experiment_results}, but with 30\% of client participation in each round.}
\label{tab:auc_results}
\end{table}

Algorithm~\ref{alg:handcrafted_detection} performs a static analysis of the model parameters $\theta$ before local training. Its complexity is independent of the client's dataset size and is instead dominated by matrix operations on each linear layer's parameters, scaling with the weight matrix dimensions, i.e., $O(n^2 d)$ per layer. This is validated empirically by the negligible overhead observed for the ResNet models. The higher overhead for the CNN is due to larger linear layers, which are the primary focus of this static analysis.
In contrast, Algorithm~\ref{alg:loss_analysis} requires computing the per-sample loss $\mathcal{L}(\cdot)$ for both the current model $f_{\theta_t}$ and the previous $f_{\theta_{t-1}}$ over the entire client dataset $\mathcal{D}_c$. This induces an overhead proportional to two full forward passes, $O(N_c \cdot \text{Cost}_{\text{fwd}})$. The empirical results in Table~\ref{tab:overhead} confirm this, showing a moderate runtime that scales with the model's depth (e.g., increasing from $1.390\,s$ for ResNet18 to $3.103\,s$ for ResNet34). 
Finally, Algorithm~\ref{alg:gradient_analysis} has the highest overhead, as theoretically predicted. It must compute per-sample gradients for both models across all $N_c$ samples, resulting in a complexity proportional to two full backward passes (which inherently include the forward passes), $O(N_c \cdot \text{Cost}_{\text{bwd}})$. This is computationally more expensive than $\text{Cost}_{\text{fwd}}$, which is clearly reflected in the empirical data, showing the largest time cost across all models.

\section{Conclusions} \label{sec:conclusions}
This paper presents a comprehensive analysis of four state-of-the-art active \glspl{gia}, with particular emphasis on their claims of stealthiness. We propose a set of lightweight detection methods that can be directly executed by \gls{fl} clients to identify such attacks. Experimental evaluations conducted across multiple configurations demonstrate that our detection techniques are highly effective in uncovering these \glspl{gia}, enabling \gls{fl} systems to promptly detect and effectively mitigate them in real-world deployments. 
Consequently, future adversaries for active \glspl{gia} face a dual challenge. First, they must embed the malicious data extraction mechanism into the architecture of the global model by avoiding architectural manipulation. Second, they must ensure the resulting manipulated model maintains behavioral consistency with the legitimate training process, thereby preserving temporal coherence with the previous trusted model.

\bibliographystyle{IEEEtran}
\bibliography{bibtex}

\section*{Ethics Considerations}
In conducting and presenting this research, we have carefully evaluated its ethical implications. We conclude that our work does not pose any ethical concerns. Our research focuses on the analysis of existing state-of-the-art active \glspl{gia} with the explicit goal of developing effective, practical, and lightweight client-side detection mechanisms. These insights are intended to guide the development of stronger defenses, mitigating and preventing the privacy risks associated with these attacks in real-world \gls{fl} deployments. All experiments are conducted using publicly available datasets, and no private or sensitive user data was involved in this study.
\appendices
\section{Background on Deep Neural Networks.}\label{sec:appendix_background}
A deep neural network \(f\) with \(N\) layers \(\{f_i\}_{i=1}^N\) maps an input \(x\in\mathcal{X}\) to an \(m\)-dimensional vector of confidence scores via the composition:
\begin{equation}
    f_{\theta}(x) \;=\; f_N\big(f_{N-1}(\dots f_1(x))\big).
\end{equation}
The parameters of the neural network, denoted as \(\theta\), are usually obtained by fitting on a labeled training dataset \(D = \{(x_i, y_i)\}_{i=1}^{|D|}\). The training process aims to minimize a predefined loss function \(\mathcal{L}\) that quantifies the difference between the model's output \(f_{\theta}(x)\) and the ground-truth label \(y\), i.e.,
\( \theta^* = \arg\min_{\theta} \sum_{(x,y)\in D} \mathcal{L}\big(f_{\theta}(x), y\big). \)

For classification problems, it is convenient to decompose \(f\) into two main components: a feature extractor \(g\) and a classifier \(h\). This decomposition enables defining the model output as: 
\(
    f_{\theta}(x) \;=\; h\big(g(x)\big).
\)
More formally, given a split index \(k \in \{1, \ldots, N-1\}\), we define:
\begin{equation}
    \begin{aligned}
    z &= g(x) = f_k\big(f_{k-1}(\dots f_1(x))\big),\\
    f_{\theta}(x) &= h(z) = f_N\big(f_{N-1}(\dots f_{k+1}(z))\big).
    \end{aligned}
\end{equation}
The feature extractor \(g\) consists of the first \(k\) layers (i.e., layers \(1\) through \(k\)) and is designed to process input data according to its characteristics. For image inputs, \(g\) can include convolution, pooling, and normalization layers, while for textual data, it often includes embedding layers followed by transformer or recurrent blocks. The role of \(g\) is to convert raw input into a meaningful \textit{latent} representation \(z \in \mathcal{Z}\), which captures relevant patterns and features.

The classifier \(h\) consists of the remaining layers (i.e., layers \(k+1\) through \(N\)) and is typically composed of one or more linear \gls{fc} layers, followed by non-linear activation functions such as ReLU. The final layer generally employs a softmax or sigmoid activation function to produce normalized class probability scores.

\begin{figure}[t]
    \centering
    \begin{adjustbox}{max width=\columnwidth}
        \includegraphics{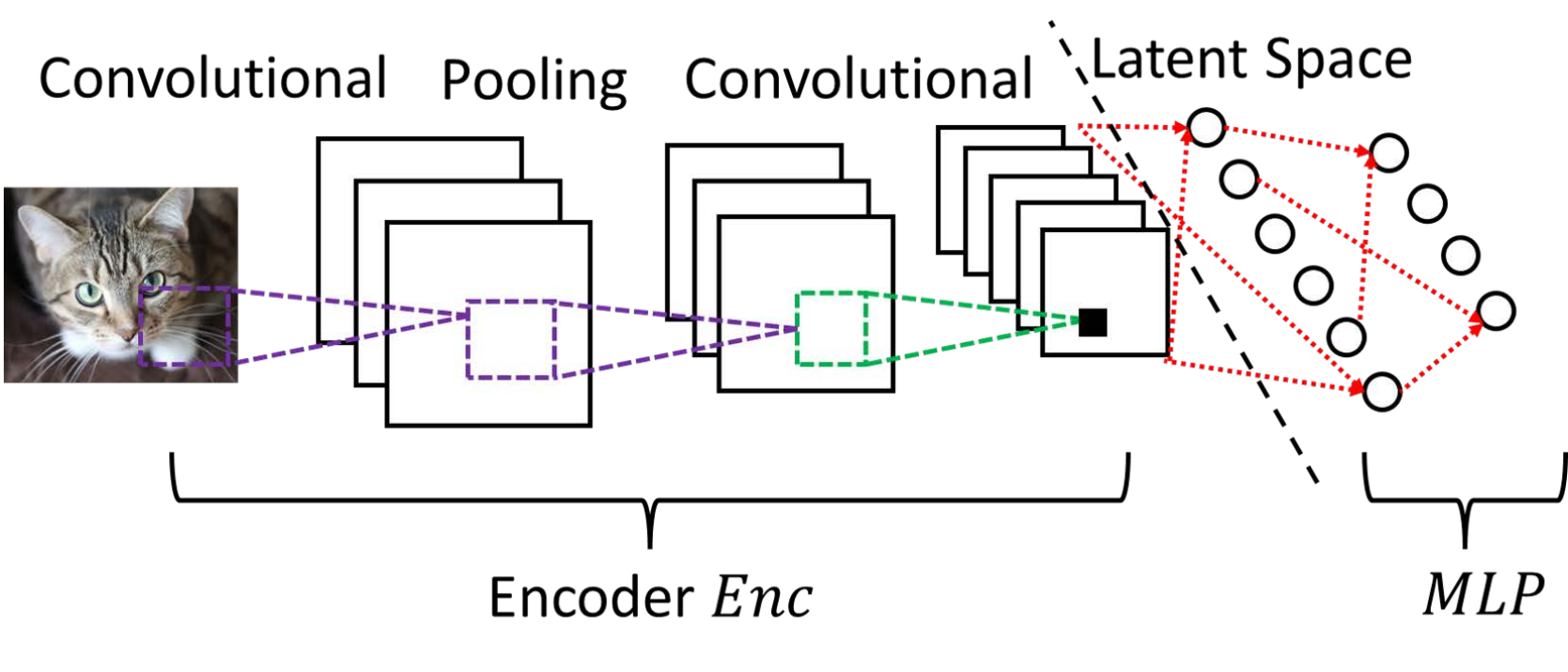}
    \end{adjustbox}
    \caption{High-level model architecture of machine-learning classifiers from Shi et al.~\cite{Shi2023ScaleMIAAS}.}    \label{fig:model_architecture_example}
\end{figure} 
\section{Binning Property-based GIAs}\label{sec:appendix}

One of the first strategy that leveraged linear layers to achieve effective data 
reconstruction is the so-called
"binning" property, introduced for the first time by Fowl et al.~\cite{rtf}, and then 
further explored by following works~\cite{loki,loki_resource,Shi2023ScaleMIAAS}.

\smallskip \noindent \textbf{Theoretical Foundation of \emph{Binning Property.}}
This technique leverages a mathematical property of neural networks with 
two subsequent linear layers and a non-linear activation function (e.g., ReLU) 
in between to \textit{perfectly reconstruct} batched inputs from aggregated gradients.
Specifically, consider a neural network composed by two consecutive linear layers defined as:
\begin{equation}
    \begin{aligned}
        out_1 &= \text{ReLU}(W_1 \cdot x + b_1) \\
        out_2 &= W_2 \cdot out_1 + b_2
    \end{aligned}
\end{equation}
where \(x \in \mathbb{R}^d\) is the input, \(out_1 \in \mathbb{R}^k\) is the intermediate representation 
(with $k$ neurons in the first linear layer), 
and \(out_2 \in \mathbb{R}^o\) is the output. The key insight is that an attacker with control over 
the model architecture (i.e., the central server in \gls{fl} as described in Section~\ref{sec:threat_model}) 
can craft these layers to enable 
perfect data reconstruction.
The attack proceeds through the following steps:

\begin{enumerate}
    \item \textbf{Linear Function Selection.} The linear leakage attack assumes that the attacker has access to an auxiliary dataset \(D_{aux}\) that 
    follows a distribution similar to those of the target inputs. This aligns with the 
    threat model defined in Section~\ref{sec:threat_model}.
    Using this dataset, the attacker select a linear function over the input data 
    space, defined by a vector \(v_{feature} \in \mathbb{R}^d\), that captures a 
    meaningful scalar feature of the data. 
    The hyphotesis is that the distribution of this feature is continuous and can be effectively estimated using the auxiliary dataset.
    Previous work use the average value of image brightness along all pixels of the image~\cite{rtf,Shi2023ScaleMIAAS,loki}. 
    The scalar feature \(feature(x) = v_{feature} \cdot x\) provides a projection of each input sample \(x\) onto this direction.

    \item \textbf{Creation of Bins.} Using the auxiliary dataset, the attacker estimates 
    the \gls{cdf} \(\psi(feature(x))\) of the feature \(feature(x)\) across the auxiliary dataset \(D_{aux}\), assumed to be continuous for the quantity measured by \(feature\). This approach does not require the server to have access to the complete distribution of user data; instead, it can estimate the distribution of a specific scalar feature derived from the user data.
    This \gls{cdf} is then used to divide the feature space into \(k\) equally probable bins by calculating the quantiles \(q_i = \psi^{-1}(i/k)\) for \(i \in \{1, 2, \ldots, k\}\). This ensures that each bin has the same probability of containing a random sample from the target distribution.

    \item \textbf{Malicious Weight Configuration.} The attacker crafts the neural network architecture to enable perfect data reconstruction by setting all row vectors of the first layer's weight matrix \(W_1\) to be identical to the feature vector \(v_{feature}\), i.e., \(W_1^{(r)} = v_{feature}\) for all \(r \in \{1, 2, \ldots, k\}\). This ensures that every neuron in the first layer computes the same scalar feature \(feature(x) = v_{feature} \cdot x\) for any input \(x\).

    \item \textbf{Bias Vector Design.} The bias vector \(b_1\) is set to the negative values of the bin boundaries: \(b_1 = [-q_1, -q_2, \ldots, -q_k]^T\). Combined with the ReLU activation function, this configuration transforms each neuron into a binary indicator: neuron \(r\) activates (outputs a positive value) if and only if the input's feature value \(feature(x)\) falls within the \(r\)-th bin \([q_{r-1}, q_r]\).

    \item \textbf{Second Layer Uniformity.} All row vectors of the second layer's weight matrix \(W_2\) are set to be identical. This ensures that the gradient with respect to each intermediate neuron is the same, i.e., \(\frac{\partial \mathcal{L}}{\partial y_i} = \frac{\partial \mathcal{L}}{\partial y_j}\) for all neurons \(i, j\) in the intermediate layer, which is crucial for the reconstruction formula.

\end{enumerate}

With this crafted architecture, when the batch size \(B\) is smaller than the number of neurons \(k\), each input sample activates exactly one neuron in the first layer, corresponding to its feature bin. This one-to-one mapping enables perfect reconstruction through gradient analysis. Specifically, if sample \(x_p\) is the only sample falling into bin \(r\), it can be perfectly recovered using:
\begin{equation}
    x_p = \frac{\nabla_{W_1^{(r+1)}} \mathcal{L} - \nabla_{W_1^{(r)}} \mathcal{L}}{\nabla_{b_1^{(r+1)}} \mathcal{L} - \nabla_{b_1^{(r)}} \mathcal{L}}
\end{equation}
where the numerator captures the difference in weight gradients between consecutive bins, and the denominator provides the necessary normalization through bias gradient differences.

% However, when \(m > k\), multiple samples may fall into the same bin, causing collisions that prevent individual sample reconstruction. In such cases, the gradients reflect a mixture of multiple inputs rather than individual samples, leading to reconstruction failures. Therefore, the number of neurons \(k\) represents a fundamental bottleneck for the linear leakage attack effectiveness.

\smallskip \noindent \textbf{Exploitation of Binning Property.}
As previously demonstrated, the binning property can be exploited to reconstruct input data with high fidelity under the assumption that the network architecture is composed of two manipulated linear layers. However, as mentioned in Section~\ref{sec:background}, real-world deep neural networks are not composed solely of linear layers, but also include convolutional layers, pooling layers, and normalization layers. Linear layers are typically employed in the classification head of the network $h$, where they transform high-level embeddings $z$ extracted by the preceding layers into final output predictions. In this context, while the gradients of the linear layers can still provide valuable information for reconstructing the embedding $z=g(x)$ produced by the feature extractor, the presence of non-linear layers before the linear layers significantly complicates the recovery of the original input data $x$.

For this reason, several works have proposed different methods to leverage the binning property in realistic network architectures. The first approach involves altering the model architecture by inserting two additional linear layers earlier in the network, typically near the input layer $f_1$~\cite{rtf}. In this way, the gradients computed by these layers can provide more direct information about the input data. This idea was further explored in subsequent works~\cite{loki,loki_resource}, where the authors placed additional linear layers before the shared model (as input layers) in combination with manipulated convolutional kernels. However, this manipulation directly alters the shared model architecture and can be easily detected by \gls{fl} participants due to the presence of additional linear layers in unconventional positions where they are not typically found in standard neural network architectures. Furthermore, since clients have prior knowledge of the model architecture, they can immediately identify such architectural anomalies.

Recently, Shi et al.~\cite{Shi2023ScaleMIAAS} proposed a different approach to exploit the binning property without altering the shared model architecture. The core idea is to leverage existing linear layers in the classification head to recover the embeddings $z$ produced by the preceding feature extraction layers $g$. Once the embeddings are reconstructed, the attacker employs a carefully crafted decoder, trained on a surrogate dataset, to map the embeddings back to the input space and recover the original input data. The authors claim that this approach achieves high-fidelity reconstruction without architectural modifications, making it more difficult to detect, particularly during the initial rounds of the \gls{fl} process~\cite{Shi2023ScaleMIAAS}.

\section{Paired Weight GIAs}\label{sec:appendix_paired}

One of the more sophisticated strategies leveraging handcrafted manipulation of linear layers to achieve effective data reconstruction was introduced by Wang et al.~\cite{mkor}. The approach employs a two-phase strategy that manipulates different components of the shared model to create a deterministic information flow path, enabling perfect reconstruction of intermediate features using an analytical formula. In the first phase, the adversary handcrafts the weights and biases of the classifier module $h$ to establish a deterministic mapping between input features and neuron activations. In the second phase, the adversary modifies the convolutional layers of the feature extractor $g$ to ensure that the reconstructed features retain sufficient spatial information for effective reconstruction of the original input data.

\begin{table*}[h!]
\centering
\small
\renewcommand{\arraystretch}{1.5}
\begin{adjustbox}{max width=\textwidth}
% Setup(l), Atk(l), Data(l), Model(l), Runs(c), TPR(c), FPR(c)
\begin{tabular}{l l l l c c c}
\toprule
  \shortstack{\textbf{FL Setup}\\\textbf{ }} & \shortstack{\textbf{Attack}\\\textbf{ }} & \shortstack{\textbf{Dataset}\\\textbf{ }} & \shortstack{\textbf{Global}\\\textbf{Model}} & \shortstack{\textbf{Number of}\\\textbf{Runs}} & \shortstack{\textbf{Average}\\\textbf{TPR$\uparrow$}} & \shortstack{\textbf{Average}\\\textbf{FPR$\downarrow$}}  \\
\midrule
    \multirow{6}{*}{\shortstack{Cross-Silo\\(10 clients)}} & \cite{garov2024hiding} & CIFAR-100~\cite{krizhevsky2009learning} & ResNet18~\cite{he2016deep} & 3 & $1.000_{\pm 0.000}$ & $0.005_{\pm 0.009}$ \\
     & \cite{garov2024hiding} & CIFAR-10~\cite{krizhevsky2009learning} & ResNet18~\cite{he2016deep} & 3 & $1.000_{\pm 0.000}$ & $0.000_{\pm 0.000}$ \\
     & \cite{garov2024hiding} & Tiny-ImageNet~\cite{deng2009imagenet} & ResNet18~\cite{he2016deep} & 3 & $1.000_{\pm 0.000}$ & $0.000_{\pm 0.000}$ \\
  \cline{2-7}
     & \cite{shan2025geminio} & CIFAR-100~\cite{krizhevsky2009learning} & ResNet34~\cite{he2016deep} & 3 & $1.000_{\pm 0.000}$ & $0.000_{\pm 0.000}$ \\
     & \cite{shan2025geminio} & CIFAR-10~\cite{krizhevsky2009learning} & ResNet34~\cite{he2016deep} & 3 & $1.000_{\pm 0.000}$ & $0.000_{\pm 0.000}$ \\
     & \cite{shan2025geminio} & Tiny-ImageNet~\cite{deng2009imagenet} & ResNet34~\cite{he2016deep} & 3 & $1.000_{\pm 0.000}$ & $0.228_{\pm 0.106}$ \\
\midrule
    \multirow{4}{*}{\shortstack{Cross-Device\\(100 clients)}} & \cite{garov2024hiding} & CIFAR-100~\cite{krizhevsky2009learning} & ResNet18~\cite{he2016deep} & 3 & $1.000_{\pm 0.000}$ & $0.017_{\pm 0.020}$ \\
     & \cite{garov2024hiding} & CIFAR-10~\cite{krizhevsky2009learning} & ResNet18~\cite{he2016deep} & 3 & $1.000_{\pm 0.000}$ & $0.004_{\pm 0.002}$ \\
  \cline{2-7}
     & \cite{shan2025geminio} & CIFAR-100~\cite{krizhevsky2009learning} & ResNet34~\cite{he2016deep} & 3 & $1.000_{\pm 0.000}$ & $0.000_{\pm 0.000}$ \\
     & \cite{shan2025geminio} & CIFAR-10~\cite{krizhevsky2009learning} & ResNet34~\cite{he2016deep} & 3 & $1.000_{\pm 0.000}$ & $0.000_{\pm 0.000}$ \\
\bottomrule
\end{tabular}
\end{adjustbox}
\renewcommand{\arraystretch}{1}
\caption{Summary of analyzed experiment configurations and detection results. All experiments simulate an IID data distribution. 
TPR/FPR values are averaged across several runs and presented as $\texttt{mean}_{\pm \texttt{std}}$. In each run, 20\% of clients are randomly selected as victims. 
Detection performance was evaluated at rounds 2, 40, 80, and 100 for the cross-silo setup, and at rounds 2, 50, 150, and 200 for the cross-device setup. The hyperparameter of the detection algorithms refers to the "Standard" configuration in Table~\ref{tab:config_parameters}.}
\label{tab:experiment_results_iid}
\end{table*}

\smallskip \noindent \textbf{Theoretical Foundation of \emph{Paired Weight} attack.}
In the first phase of the attack, the adversary handcrafts the weights and biases of the 
classifier module $h$, which can be seen as a neural network module composed by multiple linear layers defined as:
\begin{equation}
    \begin{aligned}
        z_1 &= \text{ReLU}(W_1 z + b_1) \\
        z_2 &= W_2 z_1 + b_2 \\
        &\vdots \\
        o &= W_L z_{L-1} + b_L
    \end{aligned}
\end{equation}
where $z \in \mathbb{R}^d$ is the input feature representation, $z_l \in \mathbb{R}^{n_l}$ are the 
intermediate representations, and $o \in \mathbb{R}^c$ is the final output for $c$ classes. 
The key insight is that an attacker with control over the model parameters can craft 
these layers to enable perfect feature reconstruction through coordinated parameter manipulation.
The attack proceeds through the following steps:

\begin{enumerate}
    \item \textbf{Paired Weight Configuration.} For the first fully connected layer, pairs of weight rows are configured to maintain specific mathematical relationships: $W_1[n, :] = \alpha_n W_1[n-1, :]$ for even indices $n$, where $\alpha_n < 0$ is a negative scaling factor. This configuration creates a system where exactly one neuron in each pair activates for any given input due to the ReLU activation function, establishing a deterministic mapping between inputs and neuron activations.

    \item \textbf{Coordinated Bias Setting.} The bias vector is modified to maintain consistency with the paired weight configuration: $b_1[n] = \alpha_n b_1[n-1]$ for even indices $n$. This coordination ensures that the decoupling mechanism functions correctly across different input magnitudes while preserving the mathematical relationships required for reconstruction.

    \item \textbf{Merging Path Design.} For subsequent layers, weights are configured to create deterministic merging paths while maintaining positive gradient flow: $W_l[n^{l+1}, n^l] = |W_l[n^{l+1}, n^l]|$ when specific floor-based conditions are met, creating block-structured matrices with positive-only weights in specific positions.
\end{enumerate}

With this crafted architecture, the attack enables perfect reconstruction of intermediate features using the analytical formula:
\begin{equation}
    \hat{z}_n = \sum_{k=1}^{m} \frac{\partial \mathcal{L}(x_k, y_k)}{\partial W_1[y_k, :]} \Big/ \sum_{k=1}^{m} \frac{\partial \mathcal{L}(x_k, y_k)}{\partial b_1[y_k]}
\end{equation}
where $\hat{z}_n$ represents the reconstructed feature for class $n$, and the division of weight gradients by bias gradients provides the necessary normalization.

The attack's second phase modifies the convolutional layers of the feature extractor $g$ to preserve spatial details about pixel locations and intensities. The key insight exploits the "knowledge orthogonality" principle: since natural images exhibit spatial smoothness, the attack designs convolutional filters to capture complementary information, such as maximum and minimum pixel values within spatial regions. The convolutional manipulation strategy operates as follows:

\begin{compactitem}
    \item \textbf{Dual Filter Configuration.} Standard convolutional filters are replaced with pairs of extremal value extractors. One filter in each pair extracts the sum of pixels in a region using all positive weights, while its counterpart extracts the negative sum using all negative weights, thereby providing upper and lower bound information for pixel intensities.

    \item \textbf{Systematic Spatial Coverage.} Multiple filter sets with shifted receptive fields ensure complete coverage of every input pixel by at least one filter, preventing information loss during spatial downsampling operations.

    \item \textbf{Deterministic Region Mapping.} Each feature element in the final representation corresponds to exactly one spatial region through a predetermined mapping function, eliminating ambiguity in the reconstruction process.
\end{compactitem}

The complete reconstruction process combines both phases: reconstructed features are mapped to their corresponding spatial regions, and pixel bounds are computed by identifying which filters covered each pixel location. Final pixel values are recovered by averaging the computed upper and lower bounds, enabling complete analytical reconstruction without iterative optimization.

\section{Additional Experimental Results}\label{sec:appendix_experimental}

\smallskip \noindent \textbf{Experiments with IID Data Distribution.}
Table~\ref{tab:experiment_results_iid} presents the detection performance in an IID data setting. This analysis was focused specifically on the learned \glspl{gia}~\cite{garov2024hiding, shan2025geminio}, as their corresponding detection mechanisms (e.g., loss and gradient analysis) are data-dependent. In contrast, our detection for handcrafted \glspl{gia} relies on static parameter analysis, which is independent of the client's data distribution. The results, achieving a perfect average \gls{tpr} of 1.000 and a near-zero average \gls{fpr} in almost all scenarios, are highly consistent with those observed in the more realistic non-IID setup (presented in Table~\ref{tab:experiment_results}). This consistency demonstrates that the effectiveness of our data-dependent detection algorithms is robust and not negatively influenced by the statistical data distribution among clients.

\smallskip \noindent \textbf{Detection in First FL Round.}
Our detection framework remains effective even when an attack is initiated in the very first communication round. For handcrafted \glspl{gia}, detectability is largely unaffected; these methods rely on static analysis of model parameters or architecture, which are independent of the training round~\cite{our_sok}. For learned \glspl{gia}, which are identified by behavioral divergence, the primary challenge is the absence of a previous, trusted model $\theta_{t-1}$ for comparison.

We address this by benchmarking the received model $\theta_0^\mathcal{A}$ against a legitimate \textit{baseline}. This baseline can be either a new, randomly initialized model (matching the server's claimed initialization state) or a standard, publicly available pre-trained model $\theta_{\text{pre}}$. Neither a random nor a standard pre-trained model will exhibit the highly specific, anomalous loss and gradient behaviors that the manipulated model was explicitly trained to produce~\cite{garov2024hiding,shan2025geminio}. The pronounced behavioral divergence from this legitimate baseline is sufficient to flag the attack. We have used this evaluation in partial client participation when the client is selected for the first time into an attack round, and experimental results validate our hypothesis.

However, we note that this \emph{attack-at-initiation} scenario may be less practical for an adversary. In many \gls{fl} deployments, the initial checkpoint $\theta_0$ is not an arbitrary server choice but rather a known entity, such as a client-provided checkpoint or a standard pre-trained model. In such cases, any attempt by the server to substitute this known, legitimate $\theta_0$ with a manipulated model would create a massive and immediately detectable behavioral divergence.

\begin{table}[t!]
\centering
\small
\renewcommand{\arraystretch}{1.5} % Aumenta spaziatura righe

\begin{tabular}{l r r r}
\toprule

  \multirow{2}{*}{\shortstack{\textbf{Detection}\\\textbf{Algorithm}}} & 
  \multicolumn{3}{c}{\textbf{Empirical Overhead (seconds)}} \\
  
  \cmidrule(lr){2-4}

  & \textbf{ResNet18}~\cite{he2016deep} & \textbf{ResNet34}~\cite{he2016deep} & \textbf{CNN}~\cite{Shi2023ScaleMIAAS} \\
  
\midrule
  Alg.~\ref{alg:handcrafted_detection} & $0.160_{\pm0.001}$ & $0.162_{\pm0.003}$ & $2.103_{\pm0.049}$  \\
  Alg.~\ref{alg:loss_analysis} & $1.390_{\pm0.051}$ & $3.103_{\pm0.720}$ & $13.836_{\pm1.117}$ \\
  Alg.~\ref{alg:gradient_analysis} & $10.532_{\pm0.229}$ & $18.866_{\pm1.522}$ & $16.121_{\pm1.046}$ \\
\bottomrule
\end{tabular}
\renewcommand{\arraystretch}{1}
\caption{Analysis of computational overhead for the client-side detection algorithms. Empirical runtimes are reported as $\texttt{mean}_{\pm \texttt{std}}$, averaged over 10 runs and across different clients, using the Tiny-ImageNet dataset with a local batch size of 64.}
\label{tab:overhead}
\end{table}

\smallskip \noindent \textbf{Analysis of Computational Overhead.}
We conducted an empirical evaluation of the client-side computational overhead introduced by our proposed detection methods. All experiments were performed on a system equipped with an NVIDIA A100-SXM4-80GB GPU. Table~\ref{tab:overhead} summarizes the mean and standard deviation of execution times, averaged over 10 runs using the Tiny-ImageNet dataset with a local batch size of 64. The results indicate that Alg.~\ref{alg:handcrafted_detection} (static parameter analysis) is highly efficient, incurring negligible latency (e.g., $0.160s \pm 0.001s$ for ResNet18). In contrast, the behavioral divergence methods are more computationally intensive. Alg.~\ref{alg:loss_analysis} (Loss Analysis) introduces a moderate overhead, ranging from $1.390s$ to $13.836s$ depending on the model. Alg.~\ref{alg:gradient_analysis} (Gradient Analysis) is the most resource-intensive, requiring $10.532s$ and $18.866s$ for ResNet18 and ResNet34, respectively.



% \end{thebibliography}

\end{document}